\documentclass[journal]{IEEEtran} %\documentclass[12pt,journal,compsoc]{IEEEtran}
\usepackage{float}
\usepackage{setspace}
\usepackage{amssymb,amsfonts,latexsym}
\usepackage{graphicx}
\usepackage{subfigure}
\usepackage{psfrag}
\usepackage{algorithmic}
\usepackage{algorithm}
\usepackage{comment}
\usepackage{cite}
\usepackage{url}
\usepackage{times}
\usepackage{epsfig}
\usepackage{multirow}
\usepackage{longtable}
\usepackage{amsfonts}
\usepackage{amssymb}%
\usepackage{color}
\usepackage{mathrsfs}
\usepackage{bm}
\usepackage{booktabs}
\usepackage{threeparttable}
\usepackage[cmex10]{amsmath}
\usepackage{amsmath}
\usepackage[english]{babel}
\usepackage{mathbbold}
\usepackage[table]{xcolor}
\usepackage{makecell}
\newtheorem{remark}{Remark}
\usepackage{stfloats}

\newtheorem{theorem}{\textbf{Theorem}}
\newtheorem{proof}{\textbf{Proof}}

\ifCLASSINFOpdf
\else
\fi

\begin{document}
%\doublespacing
%\title{Risk-aware Cooperative Spectrum Access for CRNs with Multiple Channels}
\title{Performance Analysis and Enhancement of  Beamforming Training in 802.11ad}

\author{\small Wen Wu,~\IEEEmembership{\small Member,~IEEE,}
%        Qinghua Shen,~\IEEEmembership{\small Student~Member,~IEEE,}    
%        Miao Wang,~\IEEEmembership{\small Member,~IEEE,}
		  Nan Cheng,~\IEEEmembership{\small Member,~IEEE,}
        Ning Zhang,~\IEEEmembership{\small Senior~Member,~IEEE,}\\
       Peng Yang,~\IEEEmembership{\small Member,~IEEE,}
          Khalid Aldubaikhy, %~\IEEEmembership{\small Student~Member,~IEEE,}
        Xuemin~(Sherman)~Shen,~\IEEEmembership{\small Fellow,~IEEE}
	
	\thanks{W. Wu, and X. Shen are with the Department of Electrical and Computer Engineering, University of Waterloo, Waterloo, ON N2L 3G1, Canada (email:\{w77wu,  kaldubai, sshen\}@uwaterloo.ca). \emph{Corresponding author: Peng Yang.}
	}
	\thanks{N. Cheng is with State Key Lab. of ISN, and with the School of Telecommunications Engineering, Xidian University, Xian 710071, China (e-mail: dr.nan.cheng@ieee.org); N. Zhang is with the Department of Computing Sciences, Texas A\&M University at Corpus Christi, TX, 78412, USA (email: ning.zhang@tamucc.edu); P. Yang is with the School of Electronic Information and Communications, Huazhong University of Science and Technology, Wuhan, 430074, P.R. China (email: yangpeng@hust.edu.cn); K. Aldubaikhy is with the Department of Electrical Engineering, Qassim University, Buraydah, 52571, Saudi Arabia (e-mail: khalid@qec.edu.sa).}}
%\thanks{Part of this work has been previously published in IEEE WiOpt 2018 \cite{wu2017Wiopt}.}

%\author{\small Ning Zhang,~
%        Ning Lu,~
%        Nan~Cheng,~
%        Jon~W.~Mark,~\IEEEmembership{\small Life Fellow,~IEEE,}
%        Xuemin~(Sherman)~Shen,~\IEEEmembership{\small Fellow,~IEEE}
%\thanks{Manuscript received September 27, 2014; revised January 28, 2015.}
%\thanks{N. Zhang, N. Cheng, N. Lu, Jon W. Mark and X. Shen are with the Department
%of Electrical and Computer Engineering, University of Waterloo, 200 University Avenue West,
%Waterloo, ON N2L 3G1, Canada (e-mail:\{n35zhang, n5cheng, n7lu, jwmark, sshen\}@uwaterloo.ca).}
%\thanks{X. Zhang is with School of Information and Software Engineering, University of Electronic Science and Technology of China, China.~
%Email: zhangx@uestc.edu.cn}}
%% <-this % stops a space}

\maketitle

\begin{abstract}
Beamforming (BF) training is crucial to establishing reliable millimeter-wave communication connections between stations (STAs) and an access point. In IEEE 802.11ad BF training protocol, all STAs contend for limited BF training opportunities, i.e., associated BF training (A-BFT) slots, which results in severe collisions and significant BF training latency, especially in dense user scenarios. In this paper, we first develop an \emph{analytical model} to evaluate the BF training protocol performance. Our analytical model accounts for various protocol components, including user density, the number of A-BFT slots, and protocol parameters, i.e., retry limit and contention window size. We then derive the average successful BF training probability, the BF training efficiency and latency. Since the derived BF training efficiency is an implicit function, to reveal the relationship between system parameters and BF training performance, we also derive an \emph{approximate expression} of BF training efficiency. Theoretical analysis indicates that the BF training efficiency degrades drastically in dense user scenarios. To address this issue, we propose an \emph{enhancement scheme} which adaptively adjusts the protocol parameters in tune with user density, to improve the BF training performance in dense user scenarios. Extensive simulations are carried out to validate the accuracy of the developed analytical model. In addition, simulation results show that the proposed enhancement scheme can improve the BF training efficiency by 35\% in dense user scenarios. %, which indicates the BF training efficiency depends on the ratio of the number of STAs to the number of A-BFT slots. Particularly, theoretical analysis indicates that the maximum BF training efficiency is $1/e$. 
%The developed analytical model can applied to evaluate the performance of other contention-based protocols in mmWave networks.

\vspace*{1mm}
\begin{flushleft}
\end{flushleft}
\vspace{-0.5cm}
%{\bf Keywords:} Cooperative spectrum access, cognitive radio networks, security, beamforming.

%-----------------------------------------------------------------------------------------------------------------------------------------------
\end{abstract}
%\newpage
%\setcounter{page}{1}
\section{Introduction}
The prevalence of emerging data-hungry services, such as myriad virtual/augmented reality gaming~\cite{abari2017enabling}, smart city\cite{wang2019smart}, social networking \cite{ota2018quoin}, and big data analytics \cite{li2018learning, li2018eccn}, is fueling the skyrocketing growth of data traffic in the near future~\cite{shen2020ai, kato2016deep}. Cisco reported  that the next five years would witness a seven-fold increase in global mobile data traffic. To tame the data tsunami, millimeter-wave (mmWave) communication emerges as the most promising wireless technology that offers a ``wire-like" connection by exploiting a swath of spectrum \cite{zhao2018beam, peng2019ai, zhao2019routing}. {Multiple standardization efforts, %, such as IEEE 802.15.3c\cite{15_3c_standard}, IEEE 802.11ad , and ECMA-387\cite{ECMA},
plethora commercial-off-the-shelf (COTS) products, and large-scale field-trials have paved the road for the success of mmWave communications. As the first ratified standard on mmWave wireless local area network (WLAN), IEEE 802.11ad can deliver a data rate of 7 Gbit/s \cite{Ad_standard}. Moreover, the on-going IEEE 802.11ay building on the legacy of 802.11ad, is anticipated to offer a data rate up to 40 Gbit/s~\cite{zhou2018ay_survey, wu2019beef, cheng2019space}.} %To further boost mmWave WLANs, Federal Communications Commission has released an extra 7 GHz unlicensed spectrum in the 64-71 GHz band for mmWave WLANs \cite{FCC}. %Moreover, the success of IEEE 802.11ad is fueling the rocket growth of commercial IEEE 802.11ad capable mobile devices in recent years. More than 600 million IEEE 802.11ad capable WiFi chip sets are envisaged to be shipped in 2020 \cite{ABISurvey}. Because of the potential market of mmWave WLAN, Federal Communications Commission (FCC) releases extra 7 GHz unlicensed spectrum in the 64-71 GHz band to double the amount of unlicensed spectrum of mmWave WLAN \cite{FCC}.  

The mmWave communications, unfortunately, suffer from severe free-space path loss because of the high-frequency band~\cite{wu2017performance}. To compensate for the path loss in mmWave communications, beamforming (BF) technology, which focuses the radio frequency power in a narrow direction, is adopted at both the transmitter and the receiver. Since reliable communication is only possible when the BF of both the transmitter and receiver are properly aligned, a \emph{BF training} procedure between transmitter and receiver is required. Without proper BF training, the data rate of mmWave WLAN may drop from several Gbps to only a few hundred Mbps \cite{zhangxinyu2016beamspy}. Hence, an efficient BF training is imperative for mmWave networks. 

Many recent works have investigated BF training schemes, such as codebook-based beam search \cite{wang2009beam}, compressed sensing schemes \cite{ marzi2016compressive}, and out-of-band solutions \cite{sur2017wifi}. Although these works can significantly improve the efficiency of BF training, they mainly consider the BF training performance from the physical layer's perspective. However, the contention feature of the BF training from the protocol's perspective is seldom considered, i.e., multiple stations (STAs) have to contend for the limited BF training opportunities, which results in severe collisions, wastes the cherished BF training opportunities and incurs extra BF training latency, especially in dense user scenarios \cite{zhou2018deep}. Hence, the elaborate analysis of the BF training protocol  is an essential process to determine the applicability of the BF training protocol in mmWave networks.

IEEE 802.11ad specifies a distributed BF training protocol. Specifically, the BF training duration is divided into several associated beamforming training (A-BFT) slots, and all STAs will distributely contend for these limited A-BFT slots in a contention and backoff manner. However, modeling the performance of BF training protocol is a challenging task. The reason is three-fold. Firstly, modeling the BF training protocol should incorporate protocol components to unveil the relationship between various protocol parameters, which requires understanding and characterizing the protocol. Secondly, due to the ``deafness" problem caused by high directionality feature of BF, i.e., an STA may not be aware of the transmission of other STAs, the BF training protocol is different from traditional carrier sensing based protocols. Hence, existing analytical models for traditional microwave WLANs are unsuitable for the BF training protocol. Thirdly, the explicit mathematical relationship between the system parameters (e.g., user density) and BF training performance is crucial to the protocol optimization in dense user scenarios. Thus, we argue that a thorough theoretical framework for 802.11ad BF training protocol is necessary and significant. Moreover, 802.11ad only provides at most eight A-BFT slots, and hence the collision probability is high in dense user scenarios, {which results in high latency, and low BF training efficiency (defined as the percentage of A-BFT slots that have been successfully utilized).} Thus, an enhancement scheme on 802.11ad is required to improve the performance in dense user scenarios.%The BFT-MAC should be context-aware, which adpatively adjusts MAC parameter in tune with the user density. 

  %Moreover, the analytical results 

In this paper, we study the performance of the BF training protocol in 802.11ad. Specifically, we target at answering the following questions: \emph{1) How good the performance of the BF training protocol is; and 2) How to further enhance  the BF training protocol performance in dense user scenarios?} Firstly, to evaluate the performance of the BF training protocol, a two-dimensional Markov chain \emph{analytical model} is developed, which takes the number of consecutive collisions and the backoff time as a state. Our analytical result unveils the impacts of the number of A-BFT slots, the number of STAs and protocol parameters, i.e., the retry limit and the contention window size, on the performance of the BF training protocol. Extensive simulation results validate the accuracy of the proposed analytical model. %Extensive simulations are carried out to validate the accuracy of the proposed analytical model. %To the best of our knowledge, this is the first analytical model for 802.11ad BFT-MAC. 
Secondly, based on the analytical model, we derive the expressions of the successful BF training probability, BF training efficiency and BF training latency. Thirdly, since the derived expression of BF training efficiency is an implicit function, to characterize the relationship between system parameters and BF training performance, we derive an \emph{approximate expression} of the BF training efficiency in dense user scenarios, which demonstrates that the BF training efficiency depends on the ratio of the number of STAs to the number of A-BFT slots. Particularly, the maximum BF training efficiency is $1/e$. Finally, since the performance substantially degrades in dense user scenarios due to the limitation of A-BFT slots in practical systems, an \emph{enhancement scheme} which adaptively adjusts protocol parameters according to the user density, is proposed to improve BF training performance. Extensive simulations are carried out and demonstrate that the enhancement scheme can significantly improve the BF training efficiency and reduce the BF training latency, as compared to 802.11ad with the default parameter setting. The main contributions in this paper are summarized as follows:
\begin{itemize}
	\item  We develop an accurate analytical model to evaluate the performance of BF training protocol. Based on the proposed analytical model, we derive the successful BF training probability, BF training efficiency and average BF training latency;
	\item We derive an approximate expression of the BF training efficiency to elaborate the relationship between system parameters and BF training performance; %Theoretical analysis indicates that the maximum BF training efficiency is $1/e$;
	\item We propose an enhancement scheme to improve the BF training performance in dense user scenarios. %Extensive simulation results demonstrate the accuracy of the proposed analytical model and the effectiveness of the enhancement scheme.
	
	%Based on the proposal model, to achieve the maximum channel utilization with the minimum cost of A-BFT slots, two protocol parameters: the number of A-BFT and retry limit, are optimized. A non-convex integer programming problem is formulated and solved to obtain the optimal parameters. Particularly, we show that the maximum channel utilization in the A-BFT stage is barely $e^{-1}$ which is the same as that of slotted ALOHA, while IEEE 802.11ad MAC with the optimal parameter setting saves about 14\% A-BFT slots compared to slotted ALOHA. Besides, analytical results indicate that the default value of retry limit in current IEEE 802.11ad systems is not optimal.
%	\item We also conduct delay analysis and find that IEEE 802.11ad MAC suffers a longer average delay compared to slotted ALOHA, which implies a trade-off exists between the cost of A-BFT slots and average delay. 
	%Average delay has been analyzed. Under the condition of maximum channel utilization, theoretical 
\end{itemize}

The remainder of this paper is organized as follows. Related work is reviewed in Section \ref{sec: related_work}. An overview of BF training procedure in 802.11ad is presented in Section \ref{sec.sys}. The proposed analytical model and the corresponding performance analysis are given in Section \ref{sec.performance_analysis}. Section \ref{sec:throughput_analysis} provides approximate analysis on the BF training efficiency. Then, Section \ref{sec.parameter setting} proposes an enhancement scheme in dense user scenarios. %Delay analysis is provided in Section \ref{sec.delay analysis}. 
In Section \ref{sec.simulation results}, extensive simulations are conducted to validate the proposed analytical model and the enhancement scheme. In Section \ref{sec.Conclusion}, concluding remarks are given, and future works are discussed. %The proofs of some theoretical results are given in the Appendix. 
%-----------------------------------------------------------------------------------------------------------------------------------------------
 \section{Related Work}\label{sec: related_work}

{A collection of works are devoted to enhancing the performance mmWave networks~\cite{yu2019millimeter, wang2016inductive, takahashi2019adaptive}. Yu \emph{et al.} proposed a novel two-stage algorithm, in which transmission clustering and path routing are jointly optimized to reduce backhaul traffic in the mmWave mesh network~\cite{yu2019millimeter}. A pioneering work developed a fast packet transmission scheduling scheme based on an inducing coloring technique, to combat the Rayleigh-fading interference in wireless networks~\cite{wang2016inductive}. Another interesting work investigated a joint optimization of beam transmit power and BF gain to adapt to the time-varying traffic in mmWave satellite communications~\cite{takahashi2019adaptive}. The above works focus on improving network performance from the perspective of resource management. As a result, they do not consider the impact of BF training on mmWave networks. In contrast, our work aims at improving mmWave network performance from the perspective of BF training.}
%An extensive body of research has been devoted to designing efficient BF training in mmWave networks  \cite{marzi2016compressive, sur2017wifi,  hassanieh2018fast}. A compressed sensing based BF training method can improve BF training efficiency by exploiting the sparse characteristic of the mmWave channel \cite{marzi2016compressive}. Some out-of-band schemes, which exploits traditional Wi-Fi signals to reduce BF training overhead \cite{sur2017wifi}. Another work proposes a fast BF training scheme by utilizing the multi-armed beam feature of directional antennas \cite{hassanieh2018fast}. While aforementioned works significantly improve the efficiency of BF training, they mainly focus on point-to-point communication systems and analyze the BF training performance from the physical layer's perspective. As a result, they do no consider the MAC layer contention when multiple STAs compete for the same BF training resource. Even with efficient BF training schemes as shown in \cite{hassanieh2018fast, marzi2016compressive}, a coarse MAC would result in severe collisions for BF training, which wastes cherish BF training resources and incurs extra BF training latency. Hence, the elaborate analysis of MAC for BF training deserves.

There are extensive research efforts on designing efficient BF training schemes \cite{wang2009beam, marzi2016compressive, sur2017wifi, hashemi2018out,hassanieh2018fast,wu2019fast,zhou2018beam}. A codebook-based search scheme is proposed in \cite{wang2009beam}, in which beamwidth is adjusted in each step until the optimal beam is identified. %To support concurrent transmission, a low-complexity hybrid BF which is a combination of analog BF and digital BF is developed in \cite{alkhateeb2015limited}. 
Marzi \emph{et al.} developed a compressed sensing based BF training method with a low complexity by exploiting the sparse characteristic of mmWave channels~\cite{marzi2016compressive}. An out-of-band scheme that exploits traditional Wi-Fi signals to reduce BF training overhead, is developed in~\cite{sur2017wifi}. Similarly, another method that exploits the spatial correlation in low-frequency band radio signals, is developed to improve BF training performance in~\cite{hashemi2018out}. In another line of research, Hassanieh \emph{et al.} proposed a fast BF training scheme by utilizing the multi-armed beam feature of directional antennas~\cite{hassanieh2018fast}. Wu \emph{et al.} developed a learning-based BF training scheme~\cite{wu2019fast}, in which the correlation structure among nearby beams is exploited to speed up BF training procedure. In addition, Zhou \emph{et al.} proposed a fast BF training scheme for the high-mobility unmanned aerial vehicles mesh networks~\cite{zhou2018beam}, which can effectively guarantee the robustness of mmWave communication links. While the aforementioned works can enhance the performance of BF training, they lack of the considerations of the contentions of multiple STAs in the BF training protocol. In contrast, our work focuses on investigating the BF training performance from the protocol's perspective.

On the other hand, some recent research works in \cite{zhou2017enhanced,shao2018two,ay_doc_A_BFT1,ay_doc_A_BFT2,ay_doc_short_SSW} are devoted to the BF training protocol. To alleviate collisions of BF training in dense user scenarios, a pioneering work \cite{zhou2017enhanced} proposed a secondary backoff scheme in the A-BFT stage, in which each STA selects not only a backoff A-BFT slot, but also a secondary backoff time within the A-BFT slot such that transmission collisions can be reduced. Leveraging the channel sparsity in the mmWave channel, Shao \emph{et al.} developed a compressed sensing method of performing BF training simultaneously for a group of STAs in \cite{shao2018two}. In addition, there are multiple industry efforts on enhancing the BF training protocol. Kim \emph{et al.} in \cite{ay_doc_A_BFT1} spread out the access attempt over time to cope with the high collision issue in dense user networks. Another draft in \cite{ay_doc_A_BFT2} allowed BF training simultaneously for different STAs over multiple channels to enhance BF training, which may increase the protocol signaling overhead. Jo \emph{et al.} proposed a short sector sweep (SSW) frame structure in \cite{ay_doc_short_SSW} which has a shorter packet length compared with an SSW frame, to increase the BF training capability in an A-BFT slot. Although these works provide efficient solutions on enhancing BF training, they more or less need to modify the protocol and hence may be incompatible with current 802.11ad. {In contrast to these works, rather than proposing new schemes with distinct features, we target on an in-depth understanding of the 802.11ad BF training protocol. The reason is two-fold. Firstly, building on the base of successful 2.4/5 GHz WiFi systems, 802.11ad is the most practical standard in mmWave WLANs, which has been widely used in many COTS mmWave devices. Secondly, the on-going 802.11ay is the follow-up of 802.11ad, which aims to enhance the legacy 802.11ad while guaranteeing backward compatibility for legacy users \cite{aldubaikhy2020low}. The 802.11ay is expected to adopt a similar BF training protocol, with an increased number of A-BFT slots \cite{ayMagazine2017}. The number of A-BFT slots is expected to be increased from 8 in 802.11ad to 40 in 802.11ay to alleviate the BF training collisions. As such, our analytical model can be readily extended to study the performance of 802.11ay. }

\begin{table}[t]
	\small

	\centering
	\caption{SUMMARY OF MAIN ACRONYMS.}
	\label{Tab:acronyms}
		\vspace{-0.1cm}
	\begin{tabular}{l l l}
		\hline
		\hline
		\textbf{Acronyms} & \textbf{Description} \\
		\hline
		A-BFT & Associated beamforming training\\
		AP & Access point\\
		ATI & Announcement transmission interval \\
		BF & Beamforming \\ 		
		BI & Beacon interval\\
		BTI & Beacon transmission interval\\
		DCF & Distributed coordinate function\\
		DTI & Data transmission interval \\
		MAC & Medium access control\\
		SSW & Sector sweep\\
		SSW-FB & Sector sweep feedback \\
		STA & Station\\
		\hline
		\hline
	\end{tabular}
\vspace{-0.35cm}
\end{table}

%-----------------------------------------------------------------------------------------------------------------------------------------------
\section{BF Training in 802.11ad}\label{sec.sys}
This section first presents the BF training procedure in 802.11ad, and then details the BF training protocol. For better understanding, Table \ref{Tab:acronyms} provides a summary of the acronyms used hereinafter.

%In this section, the network model and 802.11ad MAC for BF training are presented, respectively. Besides, as current 802.11ad can only support at most 8 STs to associate with AP in the A-BFT stage, we modify the beacon interval (BI) control frame to increase the maximum number of supported STs. 
%\begin{figure}[t]
%	\centering
%		\renewcommand{\figurename}{Fig.}
%	\includegraphics[width=0.4\textwidth]{fig/symbol_new.pdf}
%	\caption{A typical 802.11ad network. }
%	\label{fig_NetworkModel}
%\end{figure}

%\emph{Notation}: Small bold letters represent vectors;  ${\mathbb{R} ^{M \times N}}$ is the set of real matrices with $M$ rows and $N$ columns;  $\mathbb{E}\left[x\right]$ stands for the expectation of random variable $x$; $\mathbb{Z^+}$ and $\mathbb {R^+}$ are the sets of positive integers and real numbers, respectively; $C_j^i $ denotes the $i$-permutations of $j$; $\frac{\partial x}{\partial n}$ represents the partial derivative of $x$ with respect to $n$; $U\{a,b\}$ is a discrete uniform distribution in an interval $[a,b]$.
\subsection{BF Training Procedure}

We consider a WLAN network compliant with the 802.11ad standard, consisting of an access point (AP) and multiple STAs. AP coordinates the BF training, link scheduling and network synchronization in the network. Both AP and STAs adopt the directional multi-gigabit mode, i.e., each node is equipped with an electrically steerable directional antenna capable of supporting the directional transmission. %Half duplex mode is assumed for each antenna. The carrier frequency is 60GHz. 
 %Communication establishment mechanism in IEEE 802.11 ad is adopted to ensure backward compatibility. 
 
 %which consists of sector level sweep (SLS) and beam refinement protocol (BRP) \cite{Ad_standard}. In this paper, we focus on BF training of SLS. %A detailed BF training procedure is presented in the following.%In the dense network, as the number of STs is large, more beamforming training opportunities in A-BFT slots should be allocated to support these STs.

%\section{802.11ad MAC for BF Training}

\begin{figure}[t]
	\centering
%	\vspace{-0.35cm}
	\renewcommand{\figurename}{Fig.}
	\includegraphics[width=0.42\textwidth]{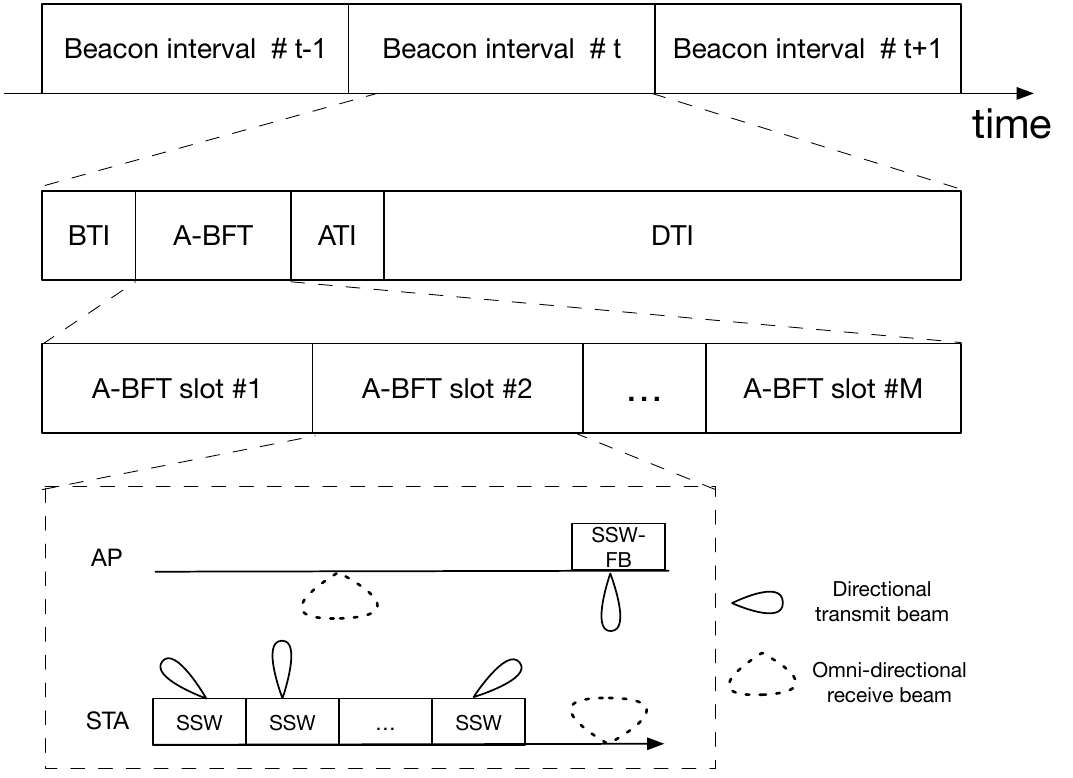}
	\caption{The 802.11ad beacon interval format and an illustration of BF training procedure.}
	\vspace{-0.35cm}
	\label{fig_BI}
\end{figure}

%\subsection{BF Training in 802.11ad}
To establish reliable mmWave communication links, an STA should perform BF training with AP at the beginning of each beacon interval (BI). As depicted in Fig. \ref{fig_BI}, the transmission time is partitioned into multiple BIs. {Each BI is further segregated into: a) beacon transmission interval (BTI), during which AP performs the BF training with STA; b) A-BFT, during which all STAs contend for BF training with AP; c) announcement transmission interval (ATI), which coordinates the transmission scheduling in data transmission interval (DTI);  and d) DTI, which facilitates directional data transmission~\cite{Ad_standard}. }%The DTI consists of multiple service periods (SPs) and contention-based access periods (CBAPs). %, where SPs are scheduled access periods and CBAPs are enhanced distributed channel access (EDCA) periods. 
Our paper focuses on the BF training in A-BFT. A-BFT is further divided into multiple A-BFT slots to provide separated BF training for different STAs. Specifically, the BF training procedure between AP and an STA in an A-BFT slot is illustrated in Fig. \ref{fig_BI}. An STA transmits multiple SSW frames via different directional beams, and AP receives these SSW frames via an omni-directional beam. Based on the received signal strength of these directional beams, AP can identify the best transmit beam of the STA. Subsequently, the AP sends an SSW-feedback (SSW-FB) frame to the transmitting STA for the acknowledgment of a \emph{successful BF training}. Note that an A-BFT slot can only provide a BF training opportunity for an STA. If two STAs compete for the same A-BFT slot, a \emph{collision} occurs. As such, no SSW-FB frame would be sent to STAs, and then the transmitting STAs are aware of the occurrence of the collision. %To facilitate the BF training for multiple STAs in the network, 802.11ad adopts BFT-MAC will be introduced in the following subsection. %Note that the BF training in this paper refers to the BF training in the A-BFT stage 

\begin{figure}[t]
%	\vspace{-0.35cm}
	\centering
	\renewcommand{\figurename}{Fig.}
	\includegraphics[width=0.54\textwidth]{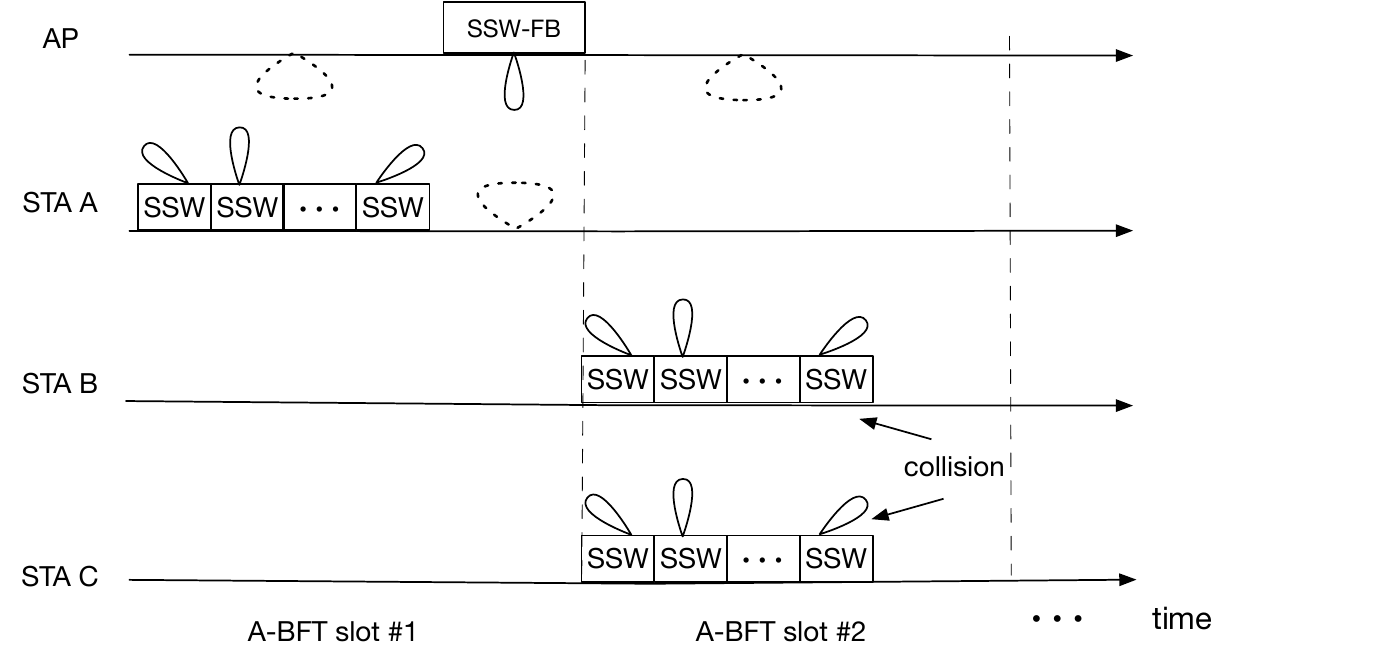}
	\caption{Illustration of the BF training protocol. The BF training in A-BFT slots \#1 is successful, while that in A-BFT slot \#2 is unsuccessful since this A-BFT slot is selected by two STAs simultaneously.}
	\label{fig_collion}
	\vspace{-0.35cm}
\end{figure}

% 802.11ad only provides at most 8 A-BFT slots (BF training resources) for all STAs in the network. Due to the limitation of A-BFT slots, the collision of BF training is severe, especially in user dense networks.  

%\begin{remark}
	Note that above mentioned BF training in A-BFT is a part of the entire BF training procedure in 802.11ad. As we focus on the BF training performance from the protocol's perspective, the detailed BF training procedure is beyond the scope of our paper. For more details, one can refer to a detailed review in~\cite{zhou2018ay_survey}.

\subsection{802.11ad BF Training Protocol}
To coordinate BF training for multiple STAs, 802.11ad specifies a distributed BF training protocol, namely, in which each STA performs BF training with AP in a contention and backoff manner. The BF training protocol consists of the following two steps.
\begin{itemize}
	\item Before BF training, each \emph{active STA} randomly selects an A-BFT slot from the range $[1,M]$ for the transmission of BF training packets (referred to as a transmission for short hereinafter). Here, $M$ denotes the number of A-BFT slots. A successful BF training depends on whether an A-BFT slot is selected by multiple STAs. If an A-BFT slot is only selected by one STA, the transmission would be successful, and an SSW-FB frame would be sent back from the AP. As shown in Fig. \ref{fig_collion}, the BF training of \emph{STA A} in A-BFT slot \#1 is successful since this A-BFT slot is not selected by any other STAs. Otherwise, if more than one STA select the same A-BFT slot, a collision would occur, and no SSW-FB would be sent back from AP, such as \emph{STA B} and \emph{STA C} in Fig. \ref{fig_collion}. 
	\item {After BF training, if the number of consecutive collisions of an STA exceeds retry limit $R$ (referred to as \emph{dot11RSSRetryLimit} in 802.11ad), the STA would select a discrete backoff time uniformly at random from $[0,W)$. }Here, $W$ denotes the contention window size (referred to as \emph{dot11RSSBackoff} in 802.11ad). Specifically, each STA maintains a consecutive collision counter (referred to as \emph{FailedRSSAttempts} in 802.11ad) which indicates the number of consecutive collisions that the STA has experienced in A-BFT. Once a collision occurs, the consecutive collision counter is incremented by one. Otherwise, upon a successful transmission, the consecutive collision counter is cleared to zero. 
\end{itemize}
%Then, 
 
Note that STAs can only transmit when the backoff time is zero. The backoff time is decremented by one after one BI. Thus, if the backoff time of an STA is $w$, the STA has to be frozen from transmission in the subsequent $w$ BIs. Due to the backoff mechanism, not all STAs are contending for A-BFT slots. We refer to the contending STAs as \emph{active STAs}, while other STAs whose backoff times are nonzero, are called  \emph{inactive STAs}. %In this paper, we assume that transmission is always successful unless a collision occurs. 

\begin{table}[t]
	\small
%	\vspace{-0.35cm}
	\centering
	\caption{Summary of notations.}
	\label{Tab:Variables_and_notations}
			\vspace{-0.1cm}
	\begin{tabular}{l l l}
		\hline
		\hline
		\textbf{Notation} & \textbf{Description} \\
		\hline
		$p$ & Conditional collision probability\\
		$p_s$ & Conditional successful transmission probability\\
		$\tau$ & The probability that an STA is active\\ 
		$C(t)$ & Consecutive collision counter at time $t$\\
		$B(t)$ & Backoff time at time $t$\\ 
		$(r,w)$ & State with $r$ consecutive collisions and $w$ backoff time\\
		$M$ & Number of A-BFT slots \\
		$R$ & Retry limit\\
		$D$ & Average BF training latency\\
		$W$ & Contention window size\\
		$S$ & BF training efficiency\\
		$\boldsymbol{\pi}$ &Steady state probability vector\\
		$T_{BI}$ & Duration of a beacon interval\\
		$T_{SSW}$ & Duration of a sector sweep frame\\
		$\mathbb{Z}^+$ & Positive integer set\\
		$F$ & Number of SSW frames in an A-BFT slot \\
		\hline
		\hline
	\end{tabular}
\vspace{-0.35cm}
\end{table}

The advantage of 802.11ad BFT training protocol is salient. Firstly, BFT training protocol is fully distributed which is scalable with the network size. Secondly, the BF training protocol is simple which can be easily implemented under different scenarios. {However, compared with the celebrated distributed coordinate function (DCF) protocol in traditional omni-directional WLANs, the absence of carrier sensing mechanism makes the BF training protocol susceptible to collision, especially in dense user scenarios. }In what follows, we analyze the performance of the BF training protocol.

\section{Analytical Model and Performance Analysis}\label{sec.performance_analysis}
In this section, we first develop a two-dimensional Markov chain analytical model for 802.11ad BF training protocol. Next, the average successful BF training probability and average BF training latency are derived. {In the following analysis, we assume a fixed number of STAs and perfect physical channel conditions (i.e., no transmission errors) with line-of-sight (LOS) communications in the network. The perfect physical channel condition assumption is widely adopted in the medium access control (MAC) protocol analysis~\cite{yang2006performance, luan2012mac, bianchi2000performance}. In perfect channel conditions, the failure of one BF training attempt can only be caused by packet transmission collision, which simplifies the MAC performance analysis since other factors (e.g., decoding error) do not need to be considered. The assumption is also reasonable in practical mmWave systems, since the 802.11ad standard is designed for indoor mmWave communications where the LOS link is dominant. As strong signal strength can be observed in short-distance LOS communications, the transmission error is negligible. Field measurements indicate an extremely low bit error rate in LOS  communications \cite{ren2017line}. Under imperfect physical channel conditions, the probability that one BF training attempt fails ($p$) is a joint probability of packet transmission collision probability ($p_c$) due to the MAC protocol and packet error probability ($p_e$) due to imperfect channel conditions, i.e., $p=1-(1-p_c)(1-p_e)$ \cite{zheng2006performance}. With the relationship, our proposed analytical model can be easily extended to the study of BF training performance under imperfect channel conditions. Hence, we have considered a perfect channel condition case in this work to better elaborate the BF training protocol performance.} In addition, we assume that each STA always needs to perform BF training at each BI~\cite{ bianchi2000performance,ye2017distributed, ye2017distributedTVT}. This assumption is reasonable in dynamic mmWave networks. Since the established communication connections are intermittent and short-lived due to the mobility and blockage, the BF training procedure would be invoked persistently. For better illustration, a summary of important notations is given in Table~\ref{Tab:Variables_and_notations}.

\subsection{Markov Model for the BF Training Protocol}
\begin{figure}[t]
	\vspace{-0.5cm}
	\centering
		\renewcommand{\figurename}{Fig.}
	\includegraphics[width=0.42\textwidth]{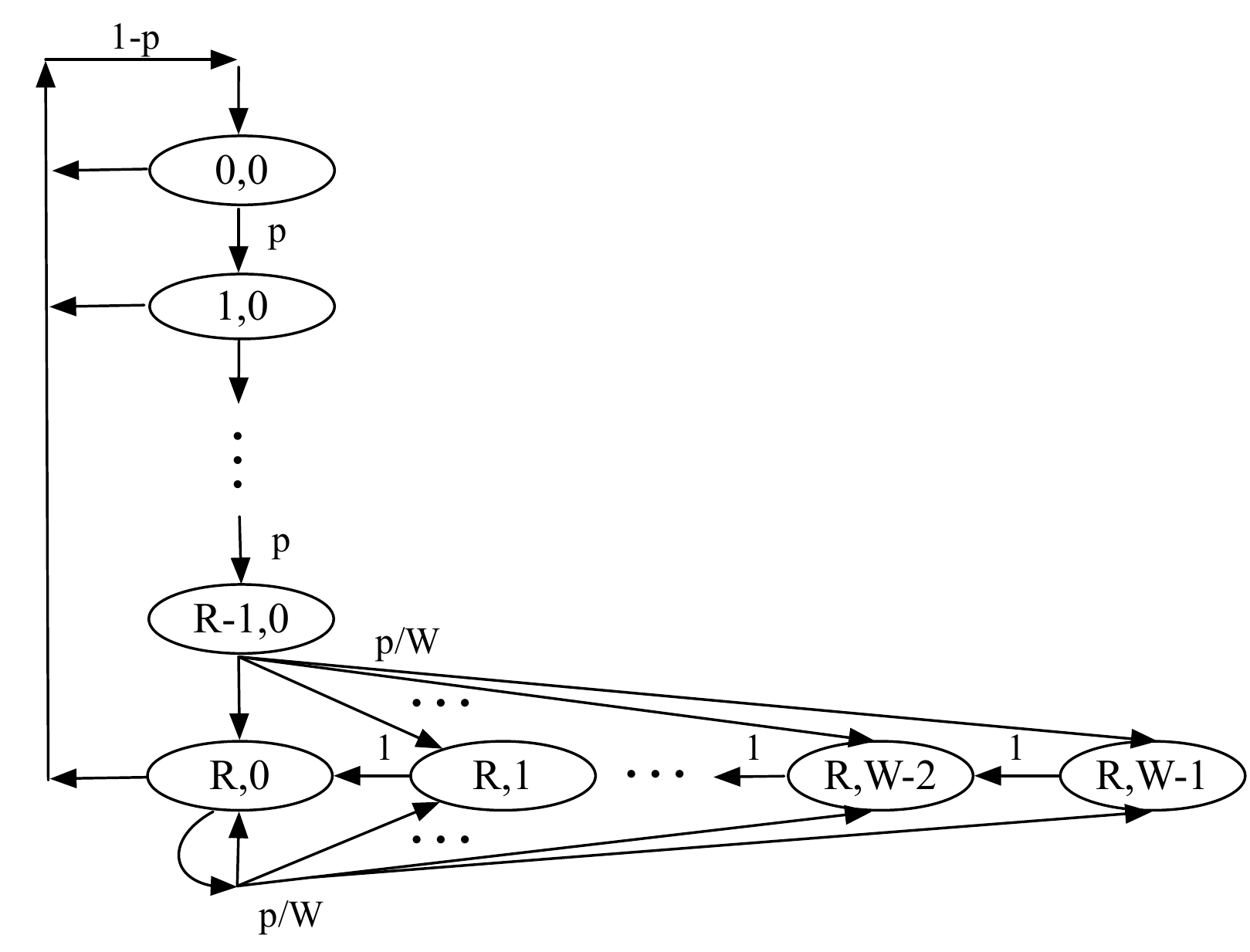}
%	\vspace{-1cm}
	\caption{Two-dimensional Markov chain model for the BF training protocol.}
	\label{fig_STD}%
	\vspace{-0.35cm}
\end{figure}

We consider a discrete time slotted system, where $t$ denotes the beginning of the $t$-th BI. To evaluate the performance of the BF training protocol, we examine a tagged STA and represent its status by a two-dimensional Markov chain $\{C(t), B(t) \}$. Let $C(t)\in[0, R]$ denote the number of consecutive collisions that the tagged STA has experienced. Let $B(t)\in[0, W-1]$ denote the current backoff time of the tagged STA. For example, {state $\left(r,w\right)$} indicates that the tagged STA has experienced $r$ consecutive collisions and its current backoff time is $w$.

Let $p$ denote the average collision probability of the tagged STA. Note that $p$ is the \emph{conditional collision probability} since the collision occurs only when the tagged STA is active. Accordingly, $1-p$ denotes the average conditional successful transmission probability. The state transition diagram is depicted in Fig. \ref{fig_STD}, which is governed by the following events and the corresponding one-step transition probabilities.

\begin{itemize}
	\item Upon a collision, the consecutive collision counter is incremented by one when it does not exceed the retry limit. The STA transits from state $(r,0)$ to state $(r+1,0)$, and the corresponding transition probability is given by
	\begin{equation}
	\label{equ:STD1Prob}
	\mathbb{P}\left(r+1,0 |r,0 \right)=p,  \forall r\in [0,R-2].
	\end{equation}

	\item Upon a successful transmission, the consecutive collision counter is cleared to zero. The STA transits from state $(r,0)$ to state $(0,0)$ with the following transition probability
	\begin{equation}
	\label{equ:STD2Prob}
		\mathbb{P}\left(0, 0 |r,0 \right)=1-p, \forall r\in [0,R].
	\end{equation}

	\item If the consecutive collision counter reaches retry limit $R$, the STA randomly selects a backoff time $w\in [0,W-1]$. The STA transits from state $(R-1,0)$ to backoff state $(R,w)$, and the corresponding the transition probability is
	\begin{equation}
	\label{equ:STD3Prob}
	\mathbb{P}\left(R, w |R-1,0 \right)=\frac{p}{W}, \forall w\in [0,W-1].
	\end{equation}
	The STA would be frozen from transmission in A-BFT in the subsequent $w$ BIs.

	\item The backoff time is decremented by one after every BI. An STA transits from state $(R,w)$ to state$(R,w-1)$, and the one step transition probability is given by
	\begin{equation}\label{equ:transition_4}
		\mathbb{P}\left(R, w-1 |R,w \right)=1, \forall w\in [1,W-1].
	\end{equation}
	
	\item The consecutive collision counter will not be incremented if it reaches the retry limit. Upon an unsuccessful transmission, an STA transits from state $(R,0)$ to backoff state $(R,w)$, and the corresponding transition probability is given by
	\begin{equation}
	\label{equ:STD4Prob0}
	\mathbb{P}\left(R, w |R,0 \right)=\frac{p}{W}, \forall w\in [0, W-1].
	\end{equation}
	Hence, the STA remains in backoff states for the subsequent $w$ BIs. 
\end{itemize}

Note that states whose backoff times are zero are called \emph{active states}, while the other states are referred to as \emph{inactive states}. Here, the steady probability of state $(r,w)$ is defined as 
 \begin{equation}
\pi_{r,w}=\lim\limits_{t\to\infty} \mathbb{P}\left(C(t)=r, B(t)=w \right)
\end{equation}
and the corresponding steady state probability of the proposed Markov chain is $\boldsymbol{\pi}=\{ \pi_{0,0}, \pi_{1,0},..., \pi_{R-1,0},\pi_{R,0},\pi_{R,1},...,\pi_{R,W-1} \}\in \mathbb{R}^{(R+W)\times 1}$. Let $\mathbf{P}\in \mathbb{R}^{(R+W)\times(R+W)}$ represent the state transition matrix whose nonnull elements are given in \eqref{equ:STD1Prob}-\eqref{equ:STD4Prob0}. Mathematically, $\boldsymbol{\pi}$ can be obtained via solving the following balance equations:
\begin{subequations}\label{equ:balance_equations}
\begin{align}
&\mathbf{P} \boldsymbol{\pi} =\boldsymbol{\pi} \label{equ:balance1} \\
&\sum_{r=0}^{R}\sum_{w=0}^{W-1}\pi_{r,w}=1\label{equ:balance2}
\end{align}
\end{subequations}
where \eqref{equ:balance2} accounts for the fact that the summation of all steady state probabilities equals one. 

\begin{remark}
	Our proposed analytical model is different from the celebrated Bianchi's model \cite{bianchi2000performance} in the following two ways. Firstly, we have different backoff mechanisms. STAs will backoff after every collision in Bianchi's model, while in our model STAs  only backoff when the consecutive collision counter exceeds the retry limit. Secondly, the contention window size increases with the number of consecutive collisions in Bianchi's model, while the contention window size is fixed in our model.
\end{remark}

%-----------------------------------------------------------------------------------------------------------------------
\subsection{Successful BF Training Probability}
%Based on the proposed Markov chain, the one step transition probabilities are given by
%\begin{equation}\label{equ: one step probability}
%	\begin{cases}
%	\mathbb{P}\left(r+1,0 |r,0 \right)=p  &\forall r\in [0, R-2]\\
%	\mathbb{P}\left(R, w |R-1,0 \right)=p/W & \forall w\in [0, W-1]\\
%\mathbb{P}\left(R, w |R,0 \right)=p/W&\forall w\in [0, W-1]\\
%	\mathbb{P}\left(0, 0 |r,0 \right)=1-p &\forall r\in [0, R]
%	\end{cases}
%\end{equation}
%where $p$ denotes the collision probability. 

Based on our analytical model, we first derive the closed-from expression of the steady state probability and then obtain the successful BF training probability.

\begin{theorem}\label{theorem_1}
The steady state probabilities of the proposed Markov chain can be represented by
\begin{eqnarray}
\pi_{r,w}= \left\{\begin{matrix}
\begin{split}
&\frac{p^r\left(1-p\right)}{p^R\left(W-1\right)/2+1}, \forall r\in[0,R-1], w = 0\\ 
&\frac{\left(W-w\right)p^R}{W\left(p^R\left(W-1\right)/2+1\right)}, \forall  r=R, w\in [0,W-1] .
\end{split}
\end{matrix}\right.
\end{eqnarray}
\end{theorem}
\begin{proof}
	The detailed proof is given in Appendix \ref{AppA}.
\end{proof}

All steady state probabilities in Theorem \ref{theorem_1} are represented by $p$ which is the unknown conditional collision probability. In the following, the value of $p$ will be obtained.

%\begin{equation}\label{equ: calculate pi_{0,0}}
%\left(1-p\right) \sum_{r=0}^{R} \pi_{r,0}=\pi_{0,0}
%\end{equation}

The probability that an STA stays in an active state, is given by
\begin{equation}\label{equ:tau}
\begin{split}
\tau=\sum_{r=0}^{R} \pi_{r,0}=\frac{1}{p^R\left(W-1\right)/2+1}.
\end{split}
\end{equation}
From the perspective of the STA, a successful transmission only occurs when other active STAs select other A-BFT slots for transmission or stay in inactive states. Hence, the conditional successful transmission probability, given an STA is active, is 
\begin{equation}
\begin{split}\label{equ: ps}
p_s&=\left(\tau\left(1-\frac{1}{M}\right)+1-\tau\right)^{N-1}\\
&=\left(1-\frac{\tau}{M}\right)^{N-1}\\
&=\left(1-\frac{1}{M\left( p^R\left(W-1\right)/2+1 \right)}\right)^{N-1}.
\end{split}
\end{equation}
The last step is obtained from the substitution of \eqref{equ:tau}. 

Since $p_s+p=1$, we can obtain the following equation:
\begin{equation}\label{equ:p_solution}
\left(1-\frac{1}{M\left( p^R\left(W-1\right)/2+1 \right)}\right)^{N-1}+p-1=0.
\end{equation}
The conditional collision probability $p$ can be obtained via solving \eqref{equ:p_solution}. However, due to the summation and permutation, Eq. \eqref{equ:p_solution} is an implicit function, and thus it is challenging to obtain a closed-form solution. Here, we apply a numerical method to obtain $p$. More importantly, Eq. \eqref{equ:p_solution} demonstrates that the conditional collision probability $p$ depends on retry limit $R$, contention window size $W$, the number of A-BFT slots $M$ and the number of STAs $N$. Obviously, these parameters are closely coupled with each other, which poses challenges on further BF training efficiency analysis.

With the obtained $p$, the successful transmission probability can be computed via the following way. Since $1-p$ represents the conditional successful transmission probability given the STA is active, the successful transmission probability is represented as follows:
\begin{equation}\label{equ:real_ps}
\hat{p}_s=\left(1-p\right) \tau
\end{equation}
which also denotes the successful BF training probability.
	
%------------------	-----------------	-----------------	-----------------	-----------------	-----------------	-----------------	-----------------	-----------------	-----------------	
\subsection{Average BF Training Latency} \label{sec.delay analysis}
In addition to the average successful BF training probability, the BF training latency is also another important performance indicator for the protocol. The average BF training latency represents the average time spent until a successful transmission. Taking the consecutive collisions before a successful transmission into account, the average BF training latency can be represented by
\begin{equation}\label{defintion_average_delay}
\begin{split}
D&=\sum_{i=0}^{\infty}\mathbb{P}\left(\text{Success}|\text{Collisions}=i\right)\mathbb{E}\left[D_i\right]\\
&=\sum_{i=0}^{\infty}\left(1-p\right)p^i\mathbb{E}\left[D_i\right]\
\end{split}
\end{equation}
where $\mathbb{E}[D_i]$ represents the BF training latency of a successful transmission after experiencing $i$ consecutive collisions. Note that if the number of consecutive collisions exceeds the retry limit, the STA would be frozen from transmission for a backoff time. Hence, the calculation of $\mathbb{E}[D_i]$ can be divided into the following two cases based on the number of consecutive collisions: 
\begin{itemize}
	\item When $i<R$, $\mathbb{E}\left[D_i\right]$ can be represented by
	\begin{equation}\label{equ:delay1}
	\begin{split}
	\mathbb{E}\left[{D}_i\right]
	&=i\cdot T_{BI}+F \cdot T_{SSW}=T_{BI}\left(i+\alpha\right)
	\end{split}
	\end{equation}
	where $\alpha={F \cdot T_{SSW}}/{T_{BI}}$. Here, $T_{BI}$ and $T_{SSW}$ denote the duration of a BI and an SSW frame, respectively. The first term in \eqref{equ:delay1} accounts for the latency caused by $i$ consecutive collisions before a successful transmission. According to the BF training protocol, if a collision occurs, the collided STA must wait until the subsequent BI to initiate a transmission attempt, which increases the latency by an entire BI duration. The second term in \eqref{equ:delay1} represents the latency for the successful transmission. As shown in Fig. \ref{fig_collion}, the successful BF training procedure consists of $F$ (referred to as \emph{FSS} field in 802.11ad) SSW frames and the corresponding BF training latency is $F \cdot T_{SSW}$. With \eqref{equ:delay1}, the average BF training latency of an STA when the number of consecutive collisions is less than $R$, is given by
	\begin{equation}\label{equ:delay_part_1}
	\begin{split}
	&\sum_{i=0}^{R-1}\left(1-p\right)p^i	\mathbb{E}\left[{D}_i\right]\\
	&=	\sum_{i=0}^{R-1}\left(1-p\right)p^i\left(i+\alpha\right)T_{BI}\\
	&=T_{BI}\left(1-p\right) \left(\sum_{i=0}^{R-1}p^i \cdot i+ \alpha \sum_{i=0}^{R-1}p^i \right)\\
	&=T_{BI}\left(\frac{ p^{R+1}\left(R-1\right)-Rp^R+p}{1-p}+\left(1-p^R\right)\alpha\right).
	\end{split}
	\end{equation}
	
	\item  When $i\geq R$, each collision results in a further random backoff time of $w$ BIs, and hence $\mathbb{E}\left[D_i\right]$ is given by
	\begin{equation} \label{equ:delay2}
	\begin{split}
	\mathbb{E}\left[{D}_i\right]
	&=((i-R+1)(\mathbb{E}\left[w\right]+1)+R-1 )T_{BI}+F \cdot T_{SSW}\\
	&=T_{BI}\left((i-R+1) (\mathbb{E}\left[w\right]+1)+R-1+\alpha\right)
  	\end{split}
	\end{equation}
	where $\mathbb{E}\left[w\right]$ denotes the average backoff time. The BF training latency when the number of a consecutive collision exceeds $R$, can be represented by
	\begin{equation}\label{equ:delay_part_2}
	\begin{split}
	&\sum_{i=R}^{\infty}\left(1-p\right)p^i\mathbb{E}\left[{D}_i\right]\\
	&=\sum_{i=R}^{\infty}T_{BI}\left(1-p\right)p^i \left((i-R+1)(\mathbb{E}\left[w\right]+1)\right.\\
	&\left.+R-1 +\alpha\right)\\
	&\stackrel{{(a)}}=\sum_{j=0}^{\infty}T_{BI}\left(1-p\right)p^{R+j}\left(\left(j+1\right)(\mathbb{E}\left[w\right]+1)\right.\\
	&\left.+R-1  +\alpha\right)\\
	&=T_{BI}\left(1-p\right)p^{R}\left(\left(\mathbb{E}
	\left[w\right]+1\right)\sum_{j=0}^{\infty}p^j \cdot j \right.	\\
	&\left.+\left((\mathbb{E}\left[w\right]+1)+R-1+\alpha\right)\sum_{j=0}^{\infty}p^j \right)\\
%	&=T_{BI}\cdot p^R\left( \frac{p(\mathbb{E}\left[w\right]+1)}{1-p}+ \mathbb{E}\left[w\right]+R+\alpha \right)\\ %+p^R\left(\left(\mathbb{E}\left[w\right]+R-1\right)T_{BI}+F \cdot T_{SSW}\right)\\
	&\stackrel{{(b)}}=T_{BI}\cdot p^R\left( \frac{W+1}{2(1-p)}+R+\alpha-1 \right)
	\end{split}
	\end{equation}
	where $(a)$ follows by changing variable $j=i-R$; $(b)$ is due to the substitution of $\mathbb{E}\left[w\right]=(W-1)/2$.% since $w$ takes a random integer within $[0,W-1]$. 
\end{itemize} 

Taking the above two cases into consideration, the average BF training latency in \eqref{defintion_average_delay} can be obtained as follows:
\begin{equation}\label{equ: delay_bound}
\begin{split}
D&=\sum_{i=0}^{R-1}\left(1-p\right)p^i	\mathbb{E}\left[{D}_i\right]+\sum_{i=R}^{\infty}\left(1-p\right)p^i\mathbb{E}\left[{D}_i\right]\\
&=T_{BI}\left(\frac{ p^{R+1}\left(R-1\right)-Rp^R+p}{1-p}+\left(1-p^R\right)\alpha\right)\\
&+T_{BI}p^R\left( \frac{W+1}{2(1-p)}+R+\alpha-1 \right)\\
&=T_{BI}\left(\frac{p^R({W-1})/{2}+p}{1-p}+a\right).
\end{split}
\end{equation}
Eq. \eqref{equ: delay_bound} demonstrates that the BF training latency depends on collision probability $p$, retry limit $R$ and contention window size $W$. Obviously, the BF training latency increases with the increase of the collision probability since severe collisions would result in substantial retransmission in the network.

\section{BF Training Efficiency Analysis}\label{sec:throughput_analysis}
%In this section, we first derive the normalized throughput and then present the asymptotic analysis on the normalized throughput in dense user scenarios.%, followed by the asymptotic analysis on the maximum normalized throughput. %To enhance the performance in dense user  scenarios, an enhancement protocol is proposed.
\subsection{Approximate  BF Training Efficiency}\label{sec: maxium normalized throughput}
%In this subsection, the normalized throughput of 802.11ad BFT-MAC is analyzed. 
Since the average successful transmission probability is $p_s\tau$, the average number of STAs that successfully perform BF training is $p_s\tau N$. The \emph{BF training efficiency} of the protocol represents the percentage of A-BFT slots that has been successfully utilized, which is defined as
\begin{equation}\label{equ:definition_S}
\begin{split}
S&=\frac{p_s\tau N}{M}\\
&=\left(1-\frac{\tau}{M}\right)^{N-1}\frac{\tau N}{M}
\end{split}
\end{equation}
where the last step follows from the substitution of \eqref{equ: ps}. Since $\tau$ depends on protocol parameters, above equation characterizes the impact of the number of STAs, the number of A-BFT slots and protocol parameters on the BF training efficiency.
 
%Substituting  into \eqref{equ:definition_S}, 
%It is very difficult to analyze the performance of the normalized throughput because $p$ is 
 From \eqref{equ:definition_S}, it is difficult to obtain the explicit relationship between protocol parameters and the BF training efficiency. This is because \eqref{equ:definition_S} is a complex function of $\tau$, $N$, and $M$. To make the performance analysis tractable, an approximate expression of the BF training efficiency is desired. According to \eqref{equ: ps}, for a large number of STAs, the conditional successful transmission probability can be approximated by
\begin{equation}\label{equ:Ps_approximation}
\begin{split}
\hat{p}_s&=\left(1-\frac{\tau}{M}\right)^{N-1}\\
&\stackrel{{(a)}}\approx\left(1-\frac{\tau}{M}\right)^{N}\\
&=\left( \left(1-\frac{\tau}{M}\right)^{{M}/{\tau}}\right)^{{N\tau}/{M}}\\
&\stackrel{{(b)}}\approx e^{-{N\tau}/{M}}
\end{split}
\end{equation}
where $(a)$ is obtained when $N$ is sufficiently large. Since we consider dense user scenarios, this condition can be easily satisfied, and hence the approximation is valid; and $(b)$ follows from the equation $\lim\limits_{n \to\infty} \left(1-{1}/{n}\right)^{n}=1/e$ where $n=M/\tau$ is sufficiently large in dense user scenarios. Here, \eqref{equ:Ps_approximation} demonstrates the average successful transmission probability is dependent on $N/M$.

Substituting \eqref{equ:Ps_approximation} into \eqref{equ:definition_S}, the \emph{approximate BF training efficiency} when the number of STAs is sufficiently large, can be given as follows:
\begin{equation}\label{equ:S_approximation}
\begin{split}
\hat{S}%&=\frac{\left(1-p\right)\tau N}{M}\\
%&\stackrel{{(a)}}= \frac{\tau N}{M} \left(1-\frac{\tau}{M}\right)^{N-1}\\
%&\stackrel{{(b)}}\approx\frac{\tau N}{M} \left(1-\frac{\tau}{M}\right)^{N}\\
%&=\frac{\tau N}{M}\left( \left(1-\frac{\tau}{M}\right)^{{M}/{\tau}}\right)^{{N\tau}/{M}}\\
&= \frac{\tau N}{M} e^{-{\tau  N}/{M}}.
\end{split}
\end{equation}
%where $(a)$ follows from $1-p=\left(1-{\tau}/{M}\right)^{N-1}$ due to \eqref{equ: ps};  

%\begin{remark}
Eq. \eqref{equ:S_approximation} characterizes the approximate BF training efficiency with respect to system parameters and provides the following important insights for protocol design in dense user scenarios. Firstly, the BF training efficiency in dense user scenarios depends on the ratio of the number of STAs to the number of A-BFT slots, i.e., $N/M$. Therefore, increasing the number of A-BFT slots adaptive to the number of STAs is an effective solution to maintain good BF training efficiency. Secondly, with a further analysis of \eqref{equ:S_approximation}, the BF training efficiency increases with ${\tau  N}/{M}$ when it is less than 1 while decreases with ${\tau  N}/{M}$ when it exceeds 1. For a practical system with a fixed number of A-BFT slots, the BF training efficiency would decrease with the increase of user density in dense user scenarios. Thirdly, 
since $\tau$ is determined by the protocol parameters, the BF training efficiency is also affected by the protocol parameters. Thus, we can conclude that the default protocol parameter setting is not always optimal in different user density scenarios, which implies that the protocol parameters should be tuned according to user density. The accuracy of this approximation is validated by simulation results in Fig. \ref{fig: throughput_vs_ratio}. 
%\end{remark}

%Therefore, intuitively increasing the number of A-BFT slots adaptive to the number of STAs is an effective solution to achieve good performance. 
\subsection{Maximum BF Training Efficiency}

%\begin{remark}

%\end{remark}

{With the approximate BF training efficiency, we target at analyzing the \emph{maximum BF training efficiency} in dense user scenarios. The maximum BF training efficiency is the maximum throughput of the BF training protocol, which provides theoretical insights on the capacity of the 802.11ad BF training protocol.} Since the number of STAs in the network is uncontrollable, the number of A-BFT slots is optimized with respect to the number of STAs to maximize the BF training efficiency. Here, we assume the number of A-BFT slots is sufficient. The problem with the limitation of A-BFT slots will be discussed in the following section. The BF training efficiency maximization problem can be formulated as follows:

%Problem I:
\begin{equation}\label{Problem first}
\begin{aligned}
{\mathcal{P}1:}	& \underset{M}{\text{max}}
& & \hat{S} \\%e^{-(1-P_{F}^T) \frac{N}{N_f}} (1-P_{F}^T) \frac{N}{N_f} \\
& \text{s.t.}
& & M\in \mathbb{Z^+}.\\
%&& & W\in \mathbb{Z^+}\\
%&& & R\in \mathbb{Z^+}
\end{aligned}
\end{equation}
The constraint indicates that the number of A-BFT slots takes positive integers. Hence, problem \eqref{Problem first} is an integer programming problem. To solve this problem, we first relax the integer constraint to a non-integer constraint. Then, this optimization problem can be readily solved by taking the derivation of \eqref{equ:S_approximation}. The corresponding condition to achieve the maximum BF training efficiency is given by
\begin{equation}\label{equ: optimal M}
M^\star=\tau N.
\end{equation} 
The condition \eqref{equ: optimal M} provides an interesting insight on the protocol design, i.e., in order to maximize the BF training efficiency, the number of A-BFT slots should equal the number of active STAs ($\tau N$) in the network. In other words, the network should provide equivalent A-BFT slots for all active users. 

Under the condition in \eqref{equ: optimal M}, the maximum BF training efficiency is
\begin{equation}\label{equ: optimal S}
\hat{S}^\star=e^{-1}.
\end{equation}

Since $\tau$ depends on the protocol parameters, the optimal number of A-BFT slots is also dependent on the protocol parameters. When the condition in \eqref{equ: optimal M} is satisfied, we have $p=1-1/e$ because of \eqref{equ:Ps_approximation}. Then, according to \eqref{equ:tau}, \eqref{equ: optimal M} can be rewritten as
\begin{equation}\label{equ:M^star_new}
M^\star=\frac{N}{\left(1-e^{-1}\right)^R\left(W-1\right)/2+1}.
\end{equation}
which characterizes the relationship between the optimal number of A-BFT slots and protocol parameters ($R$ and $W$). With a further analysis of \eqref{equ:M^star_new}, the optimal number of A-BFT slots decreases with the decrease of $R$ and the increase of $W$. To achieve the maximum BF training efficiency with limited A-BFT slots, we should choose a small value of $R$ and a large value of $W$. The detailed optimization of protocol parameters is given in Section \ref{sec.parameter setting}.

%When the condition is satisfied, the maximum normalized throughput is give $e^{-1}$.

\begin{remark}
	Approximate analysis of the maximum BF training efficiency reveals two useful insights into the performance of the BF training protocol. Firstly, %the average successful transmission probability depends on the ratio between the number STAs and the number of A-BFT slots, i.e., $N/M$, as shown in \eqref{equ:Ps_approximation}. The average successful transmission probability increases with the ratio, which demonstrates that increasing BF training resources can relieve the collision. Secondly, 
	the maximum BF training efficiency is only $1/e$. The BF training efficiency is low because of severe collisions in BF training procedure. Secondly, to achieve the maximum BF training efficiency, the optimal number of A-BFT slots should equal the number of active STAs in the network. A mismatch between the active STAs and A-BFT slots leads to performance degradation. % Finally, analytical results indicate that the normalized throughput also depends on the protocol parameters $R$ and $W$. 
\end{remark}

%-------------------- ---------------------------------------- -------------------- --------------------
\section{Enhancement Scheme}\label{sec.parameter setting}
%Simulation results indicate that the normalized throughput will decrease in dense networks. As $N$ and $M$ are not directly controlled variables, we aim to tune the protocol parameters $R$ and $W$ with the user density in the network to enhance MAC  performance.
%In other words, the provided BF training resource is limited. How to support 
In this section, we propose an enhancement scheme to improve BF training performance in dense user scenarios. 
%Note that , and the MAC parameters ($R$ and $W$). 
Previous analysis indicates that the BF training efficiency depends on the ratio of the number of STAs to the number of A-BFT slots. However, the number of A-BFT slots in current 802.11ad is limited. Specifically, 802.11ad only provides at most 8 A-BFT slots for BF training. Even though future 802.11ay increases the number of A-BFT slots to 40, the number of A-BFT slots is still limited \cite{ayMagazine2017}. The limitation of A-BFT slots renders the performance degradation in dense user scenarios.

To improve the BF training efficiency in dense user networks, we propose an \emph{enhancement scheme} that protocol parameters ($R$ and $W$) should be tuned with the user density in the network. The reasoning behind the enhancement scheme is that adjusting protocol parameters can determine the probability that STAs are active ($\tau$), such that the number of active STAs could be equivalent to the number of provided A-BFT slots. This reason is also validated by our analytical results. Since $\tau={1}/{(p^R\left(W-1\right)/2+1)}$ in \eqref{equ:tau}, $\tau$ decreases with the decrease of $R$, i.e., a small value of the retry limit renders STAs prone to enter backoff states and thus reduces the number of active STAs in the network. In addition, $\tau$ also decreases  with the decrease of $W$, because large contention window size makes STAs keep inactive for a longer time. Therefore, adaptively adjusting protocol parameters in tune with user density is an effective solution in dense user scenarios.%a small value of the retry limit and a large value of the contention window should be configured in dense user scenarios. 

%As a result, severe collision issue can be alleviated and the normalized throughput can be enhanced.

%Based on the analytical model, to achieve the best performance, 

%In 802.11ad systems, the protocol parameters usually adopt fixed default values which may not always be optimal. 
The proposed enhancement scheme mainly consists of two steps. Firstly, AP obtains the number of STAs in the network. Secondly, based on the number of STAs in the network, AP configures the optimal parameter setting of the BF training protocol and broadcasts the protocol parameters to all STAs in the network. The key issue of the enhancement scheme is the optimal protocol parameters, which could be obtained by solving the following optimization problem with the objective to maximize the BF training efficiency:% via optimizing the retry limit and the contention window:

\begin{subequations}\label{Problem 1}
\begin{align}
{\mathcal{P}2:}	& \underset{R,W}{\text{max}}
& & S \nonumber \\%e^{-(1-P_{F}^T) \frac{N}{N_f}} (1-P_{F}^T) \frac{N}{N_f} \\
& \text{s.t.}
%& & M\in \mathbb{Z^+}\\
%& & \\
%& &&\\
& & 1\leq W\leq W_{max} \label{constrain_W_max},W\in \mathbb{Z^+} \\
& &&1\leq R\leq R_{max}\label{constrain_R_max}, R\in \mathbb{Z^+}
\end{align}
\end{subequations}
where $W_{max}$ denotes the maximum contention window size, which is applied to avoid infinite backoff time. Similarly, $R_{max}$ denotes the maximum value of the retry limit. 
%\begin{subequations}\label{problem1}
%	%	\small
%	\begin{align}
%	{\mathcal{P}2:}	& \underset{\pi \in {\Pi}}{\text{min}}
%	& & R^{\pi}(T) \nonumber \\
%	& \text{s.t.}  
%	& &  \sum_{b_i\in \mathcal{B}}^{} N_{b_i}^{\pi}(T)\leq  T\\ \label{problem1, constaint 1}
%	& &&N_{b_i}^{\pi}(T)\in \mathbb{Z}, \forall b_i\in \mathcal{B}. %\\\label{problem1, constaint 2}
%	%	& &&P_{k,m}\leq P_{max}, \forall k,m\\
%	%	&&& \text{SINR}_{k,m}	\geq S_{k,m}\gamma, \forall k,m
%	\end{align}
%\end{subequations}
%The objective function is to maximize the normalized throughput.  Constraints \eqref{constrain_W_max} and \eqref{constrain_W} indicate that the contention window takes a positive integer value within the range $[1,W_{max}]$. 

%Hence, because of  and
	 The problem is challenging to be solved due to the following two reasons. Firstly, the optimization variables obey integer constraints. Secondly, the objective function is non-convex. This is because $S$ is an implicit function with respect to $R$ and $W$, since two variables $R$ and $W$ are coupled according to \eqref{equ:p_solution}. {Therefore, problem $\mathcal{P}2$ is an integer non-convex problem, which is difficult to obtain the optimal solution. However, because the number of possible combinations of the protocol parameters is limited, we can solve this problem via an exhaustive search method \cite{luan2012mac}. Since $W$ takes positive integer values within $(0, W_{max}]$, the number of possible values that $W$ can take is $W_{max}$. Similarly, for integer variable $R\in(0, R_{max}]$, the number of possible values that $R$ can take is $R_{max}$. In this way, the number of possible combinations of $R$ and $W$ is $R_{max}W_{max}$. To obtain the optimal value, the exhaustive search method needs to enumerate $R_{max}W_{max}$ combinations. Hence, the computational complexity of the exhaustive search method is $O(W_{max}R_{max})$.} In addition, to reduce computational time, the optimal protocol parameters can be computed offline with different numbers of STAs and loaded into AP as a table. Then, based on the number of STAs in the coverage of AP, the AP could search the table to obtain the optimal protocol parameters with low complexity.

%-----------------------------------------------------------------------------------------------------------------------------------------------

\section{Simulation Results}\label{sec.simulation results}
%Simulation results are presented to evaluate the proposed model and analytical results of  802.11ad MAC. 
In this section, we evaluate the proposed analytical model and the enhancement scheme via extensive Monte-Carlo simulations.%, and then compare with default 802.11ad and the celebrated slotted ALOHA protocols. % The simulation setup is first presented in subsection \ref{subsec: simulation_setup}
  \begin{table}[t]
	\small
	\centering
	\caption{Simulation parameters \cite{Ad_standard, ay_doc_short_SSW}.}
	\label{Simulation parameters}
			\vspace{-0.1cm}
	\begin{tabular}{l l l l}
		\hline
		\hline
		\textbf{Parameter} & \textbf{Value} \\
		\hline
		BI duration ($T_{BI}$)&100  \emph{ms}\\
		SSW frame duration ($T_{SSW}$)& 15.8 \emph{us} \\
		Number of STAs ($N$) &  [8, 32]\\
		Number of A-BFT slots ($M$) & 8\\
		Retry limit ($R$) & 8\\
		Contention window size ($W$) & 8\\
		Number of SSW frames in an A-BFT slot $(F)$ & 16\\
		Frequency band & $ 60$ GHz\\
		Maximum retry limit ($R_{max}$) & 20\\
		Maximum contention window size ($W_{max}$) & 20\\
		\hline
%		\hline
	\end{tabular}
\vspace{-0.35cm}
\end{table}

\subsection{Simulation Setup}\label{subsec: simulation_setup}

We validate the proposed analytical model based on a discrete event simulator coded in Matlab. We consider an 802.11ad system which operates in the unlicensed 60 GHz band. Specifically, we simulate 10,000 BIs and study the statistics of interests. {The BI duration is set to 100 ms according to a typical value in practical 802.11ad systems. Unless otherwise specified, we set $M=8$, $W=8$ and $R=8$ based on the default configuration of the 802.11ad standard~\cite{Ad_standard}. Other important simulation parameters are listed in Table \ref{Simulation parameters}.} For each experiment, we conduct 1000 simulation runs and plot a 95 percent confidence interval for each simulation point. 

%The number of users in the network is chosen from 1 to 32. The number of A-BFT slots is chosen from 12 to 20. The number of retry limit is set to be 4 unless specified. 
\begin{figure}[t]
	\centering
	\renewcommand{\figurename}{Fig.}
	\includegraphics[width=0.4\textwidth]{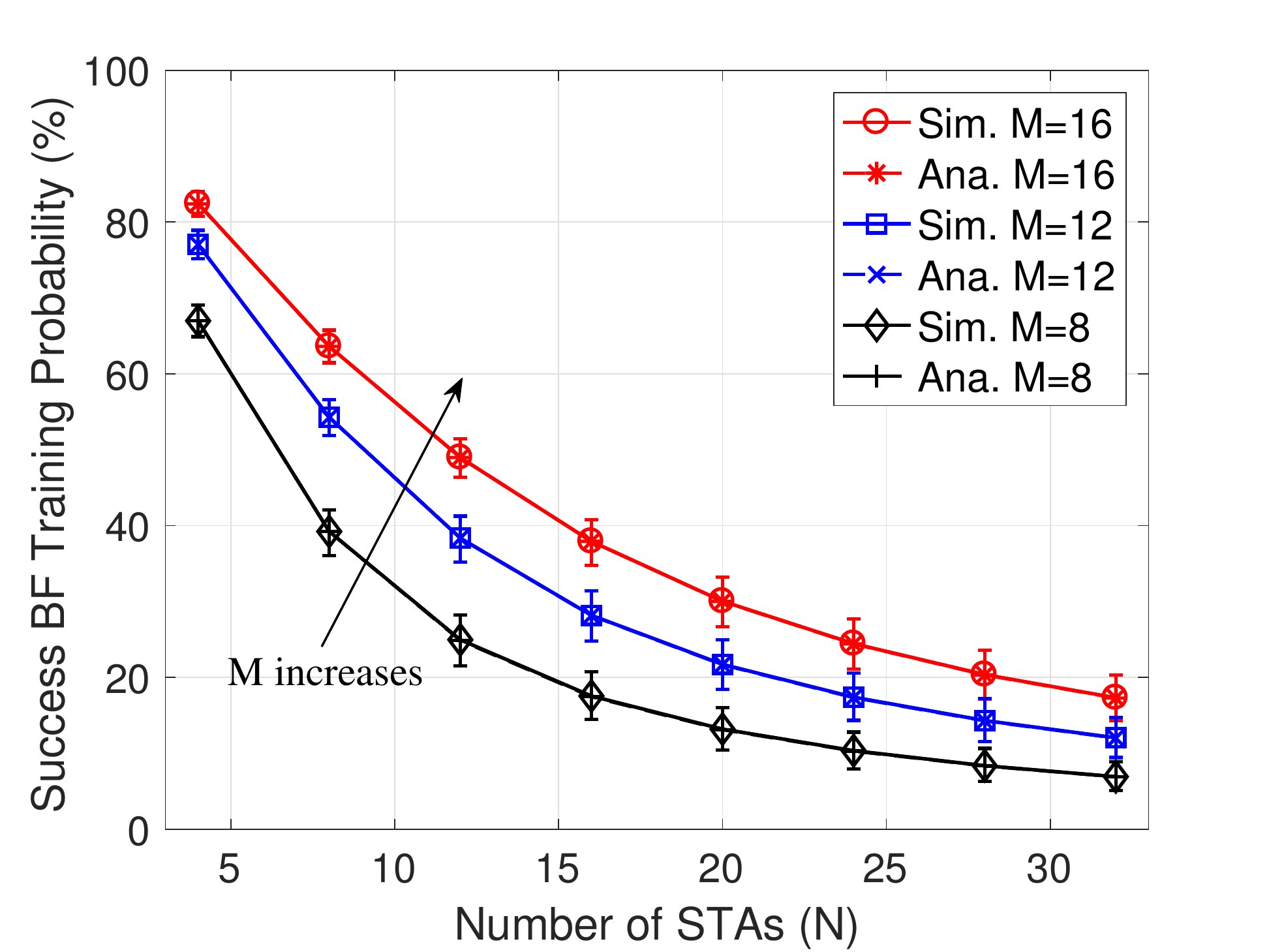}
	\caption{Successful BF training probability in terms of the number of STAs.}
	\label{AnalysisEffectiveness}
	\vspace{-0.35cm}
\end{figure}

\begin{figure}[t]
	\centering
	\renewcommand{\figurename}{Fig.}
	\includegraphics[width=0.4\textwidth]{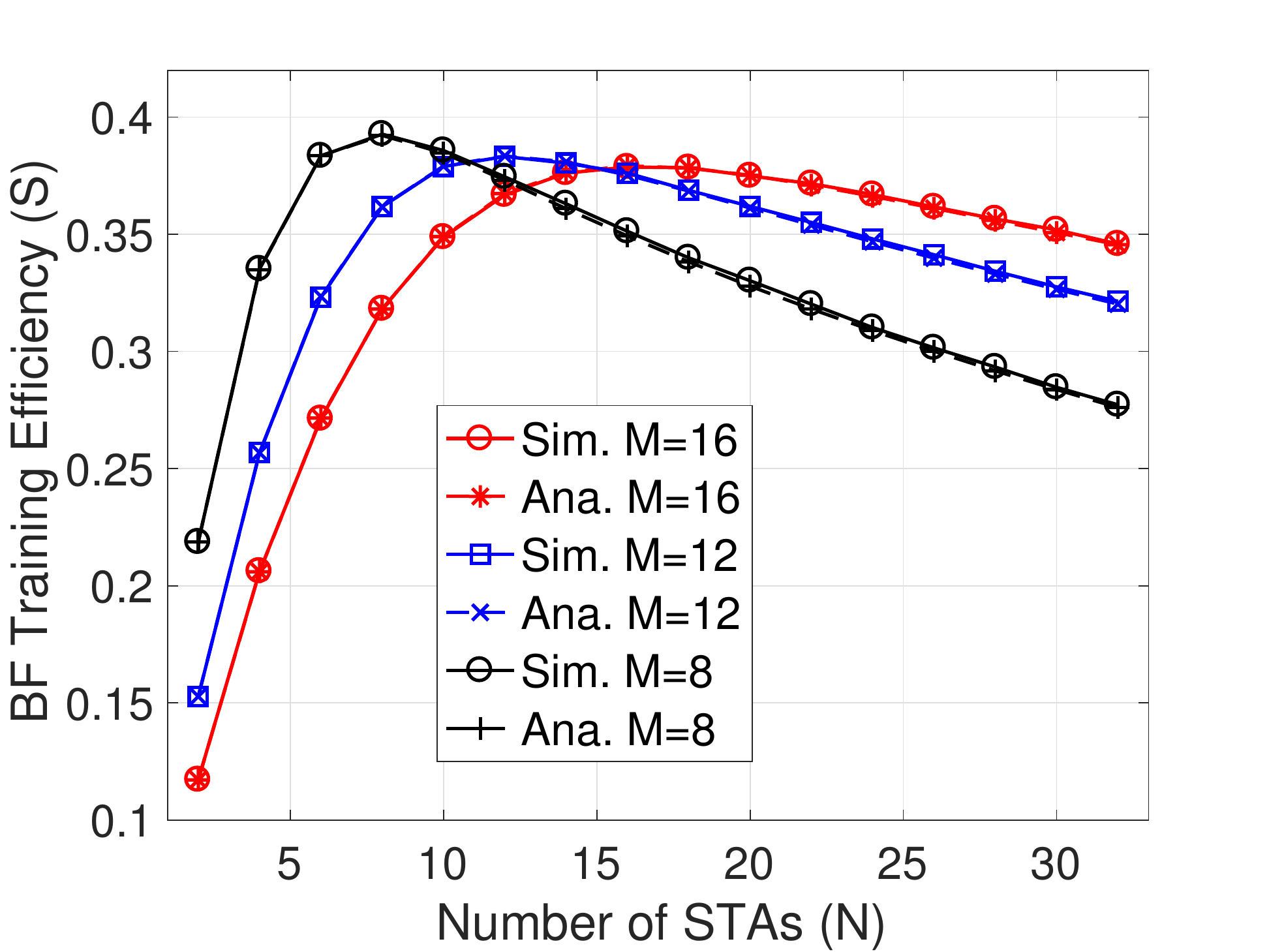}
	\caption{BF training efficiency with respect to the number of STAs.}
	\label{average_success_validate}
	\vspace{-0.35cm}
\end{figure}

\begin{figure}[t]
	\centering
	\renewcommand{\figurename}{Fig.}
	\includegraphics[width=0.4\textwidth]{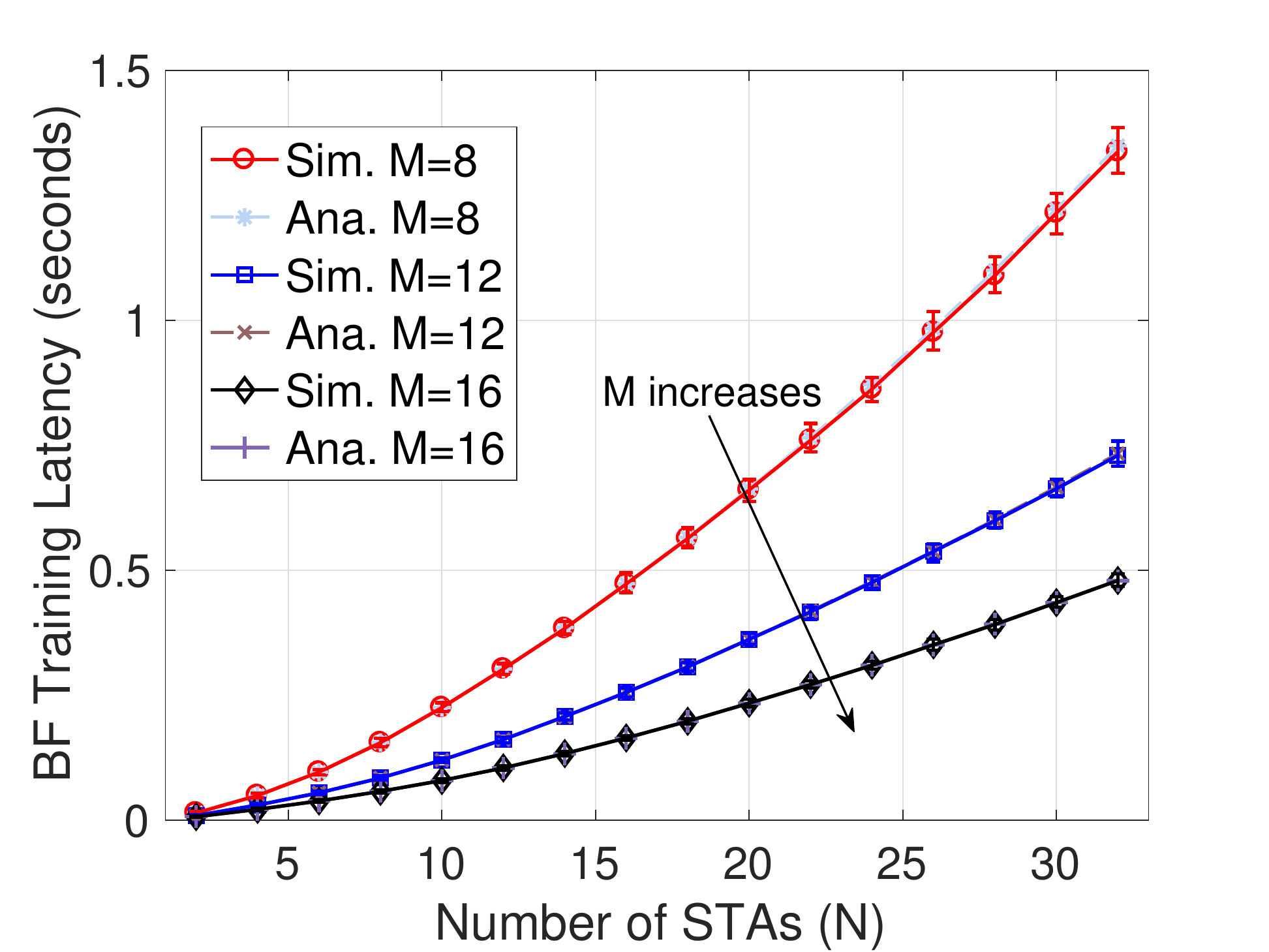}
	\caption{Average BF training latency with different numbers of STAs.}
	\label{average_delay_validate}
	\vspace{-0.35cm}
\end{figure}

\begin{figure}[t]
	\centering
	\renewcommand{\figurename}{Fig.}
	\includegraphics[width=0.4\textwidth]{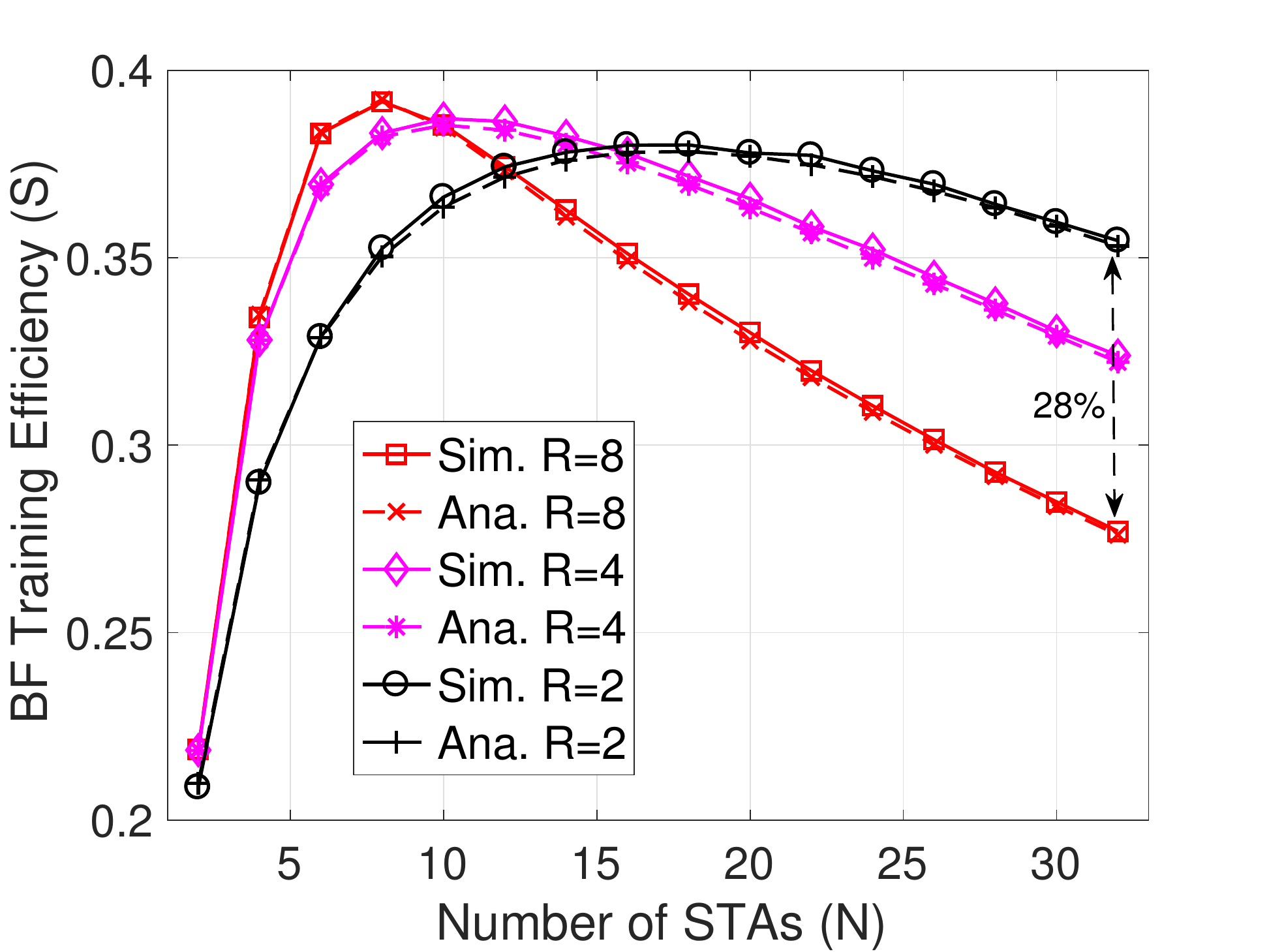}
	\caption{BF training efficiency with different values of the retry limit.}
	\label{fig: throughput_validate_vs_R}
	\vspace{-0.35cm}
\end{figure}
\begin{figure}[t]
	\centering
	\renewcommand{\figurename}{Fig.}
	\includegraphics[width=0.4\textwidth]{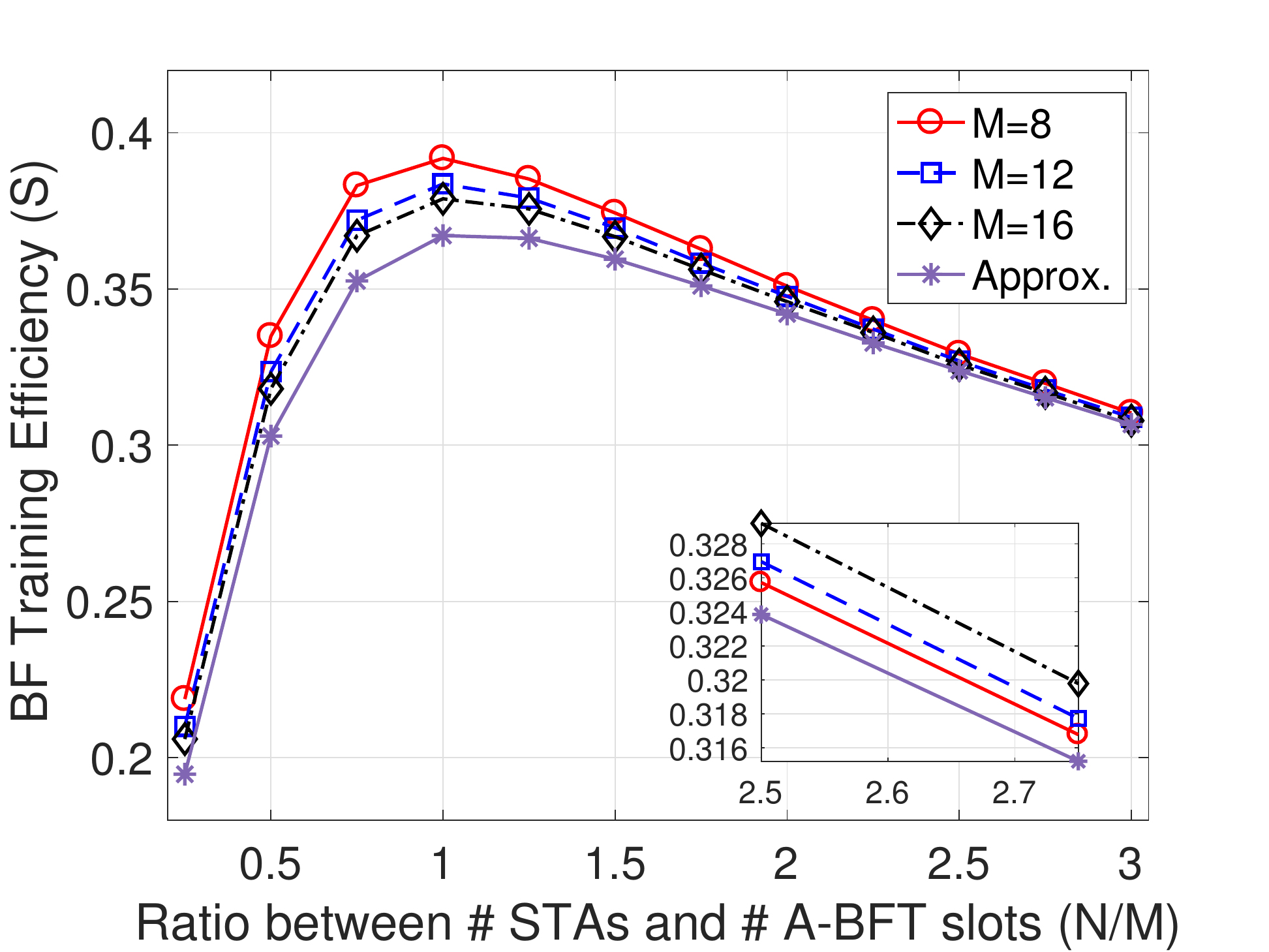}
	\caption{BF training efficiency with respect to different ratios of the number of STAs to the number of A-BFT slots.}
	\label{fig: throughput_vs_ratio}
	\vspace{-0.35cm}
\end{figure}

\subsection{Analytical Model Validation}
	Figure \ref{AnalysisEffectiveness} shows the successful BF training probability in terms of the number of STAs for $M=8, 12,16$. It can be seen that the results obtained via our analytical model are highly consistent with that via simulations. As expected, we can see a lower successful BF training probability in denser user scenarios since more STAs contend for limited A-BFT slots. The successful BF training probability drops from more than 80\% to less than 20\% as the number of STAs increases from 4 to 32. Moreover, as the number of A-BFT slots ($M$) increases, the successful BF training probability increases because more A-BFT slots are provided. The results validate that the successful BF training probability mainly depends on the numbers of STAs and A-BFT slots. {Note that the analytical results shown in Figs. \ref{AnalysisEffectiveness}-\ref{fig: throughput_validate_vs_R} are obtained via a numerical solution of (11), instead of the approximation in (20) which is valid only for a large number of STAs. Hence, the shown results are accurate even for a small number of STAs.}

In Fig. \ref{average_success_validate}, we further show the BF training efficiency with respect to the number of STAs. Several important observations can be made. Firstly, simulation results are closely matched with analytical results, which further validates our analytical model. Secondly, the BF training efficiency exhibits a bell-shape behavior. The reason is two-fold: a) many A-BFT slots are not utilized in low user density scenarios; and b) severe collision occurs in high user density scenarios, which results in a low BF training efficiency. Thus, to achieve the maximum BF training efficiency, the number of A-BFT slots should be cautiously selected. Thirdly, a system with more A-BFT slots achieves a higher BF training efficiency than that with fewer A-BFT slots. For example, in a dense user scenario ($N=32$), the BF training efficiency for $M=16$ is 25\% more than that for $M=8$. This is because more A-BFT slots can effectively alleviate the collision issue in dense user scenarios. Finally, we can observe the maximum BF training efficiency is around $1/e$ (0.37), which complies with our analytical results.

Figure \ref{average_delay_validate} shows the impact of the number of STAs on the average BF training latency. It can be observed that simulation results comply with our analytical results in \eqref{equ: delay_bound}. Clearly, the BF training latency increases with the number of STAs. This is due to a large amount of retransmission of STAs that suffer from severe collisions in dense user scenarios. Moreover, the average BF training latency decreases as the number of A-BFT slots increases.  Specifically, when $N=32$, the average BF training latency for $M=8$ is up to 1.3 seconds, which is 150\% more than that for $M=16$. Hence, increasing the number of A-BFT slots can effectively reduce the BF training latency. %However, the number of A-BFT slots is limited in 802.11ad.

Figure \ref{fig: throughput_validate_vs_R} plots the impact of the value of the retry limit on the BF training efficiency. %We first observe that the results of the proposed analytical model match that of the practical 802.11ad system for different values of $R$. In addition, t
%The BF training efficiency varies with different values of retry limit. 
Specifically, in dense user scenarios, the protocol with a small value of the retry limit achieves a higher BF training efficiency than that with a large value of retry limit. For example, when $N=32$, the protocol for $R=2$ achieves around 28\% performance gain as compared to that for $R=8$. The reason is that a small value of the retry limit renders STAs susceptible to enter backoff states. As such, fewer active STAs will contend for A-BFT slots, thereby enhancing BF training efficiency in dense user scenarios. More importantly, since $R=8$ is the default configuration of the retry limit, we claim that the default protocol configuration is not optimal in dense user scenarios. Adaptively adjusting protocol parameters in tune with the user density is a potential solution to improve BF training efficiency.

As plotted in Fig. \ref{fig: throughput_vs_ratio}, we show the BF training efficiency in terms of the ratio of the number of STAs to the number of A-BFT slots. The BF training efficiency with the same ratio for different numbers of A-BFT slots and STAs are quite close, which complies with our analytical result that the BF training efficiency mainly depends on the ratio of contending STAs to the provided A-BFT slots. In addition, as $M$ increases, the gap between simulation results and our approximation in \eqref{equ:S_approximation} narrows. The gap can be ignored in dense user scenarios, i.e., the ratio is larger than 2, which validates the accuracy of our approximation. 

 %Thus, it is very challenging to support large numbers of STs in dense networks.

%In summary, simulation results are highly consistent with our analytical results with varying system parameters. Moreover, the performance of BFT-MAC is keenly dependent on the retry limit, the number of STAs and A-BFT slots. Adjusting the value of retry limit on the basis of the user density in the network can significantly enhance MAC performance.

\begin{figure}[t]
	\centering
	\renewcommand{\figurename}{Fig.}
	\begin{subfigure}[M=8]{
			\label{fig:enhancedProtocol_M=8}
\includegraphics[width=0.4\textwidth]{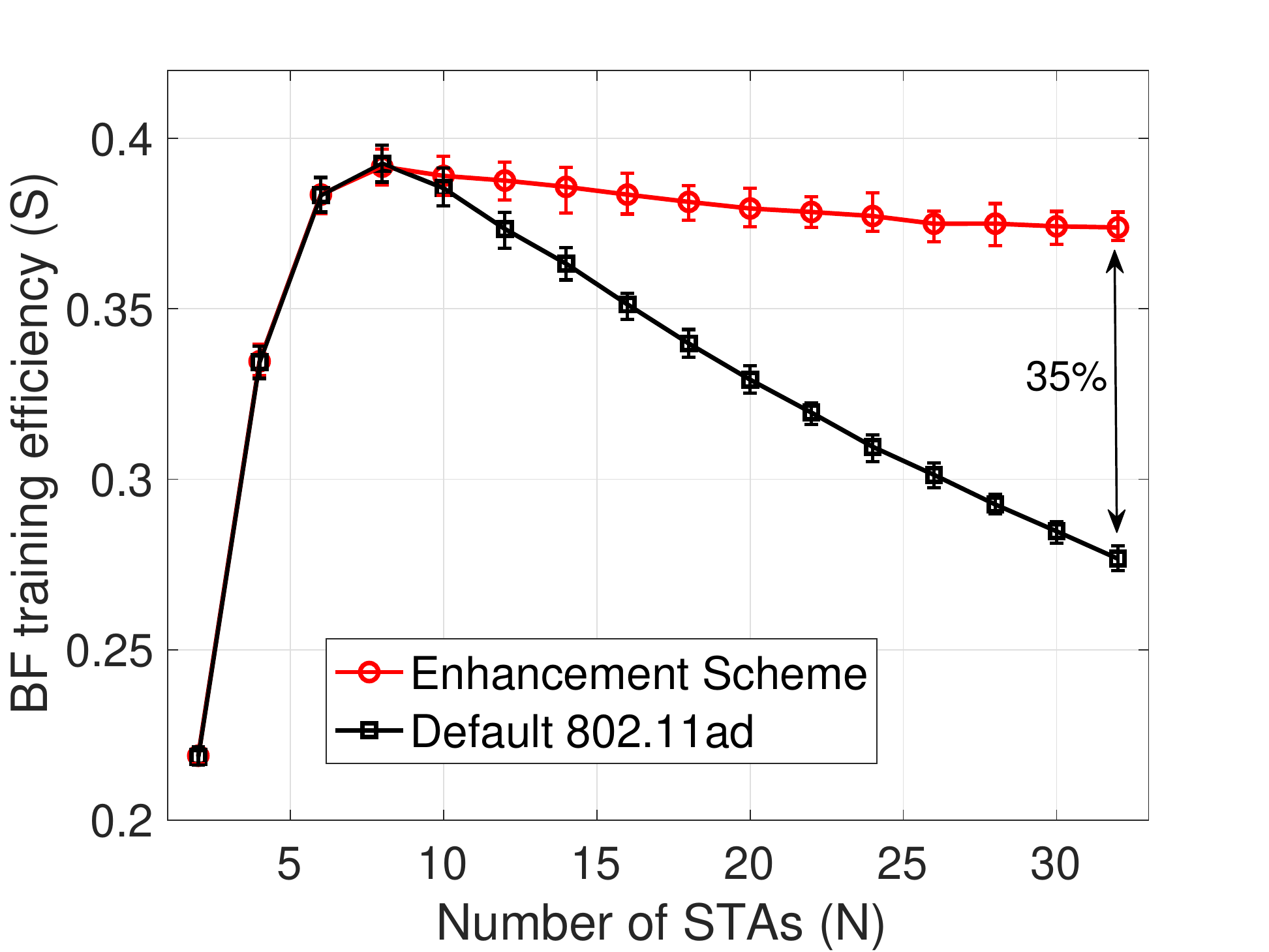}}
	\end{subfigure}%
	~
	\begin{subfigure}[M=12]{
			\label{fig:enhancedProtocol_M=12}
			\includegraphics[width=0.4\textwidth]{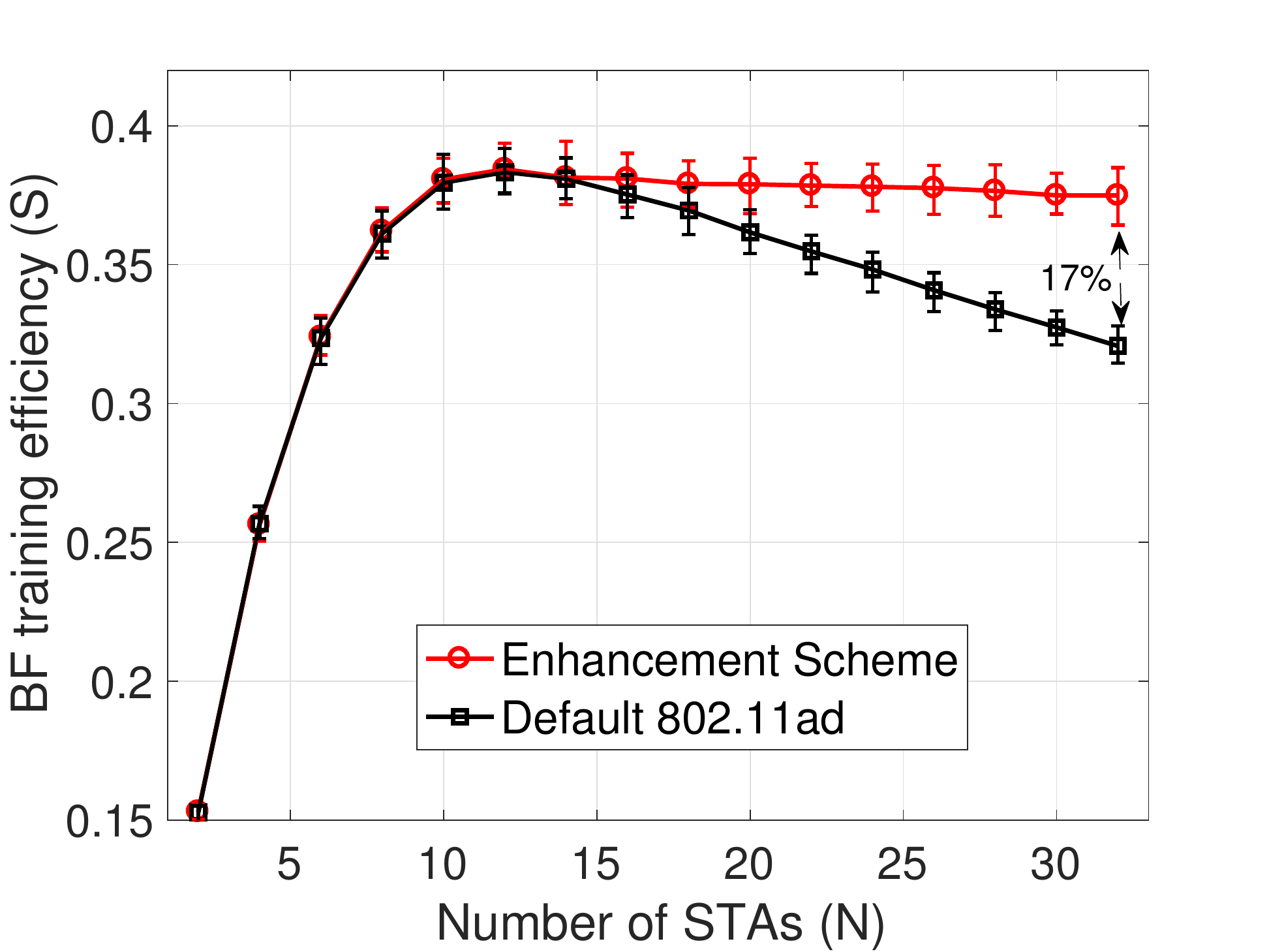}}
	\end{subfigure}
	\caption{BF training efficiency comparison in terms of different system parameters.}
	\label{Fig:enhancedProtocol}
	\vspace{-0.35cm}
\end{figure}

\begin{figure}[t]
	\centering
	\renewcommand{\figurename}{Fig.}
	\begin{subfigure}[M=8]{
			\label{fig:enhanced_delay_M=8}
			\includegraphics[width=0.4\textwidth]{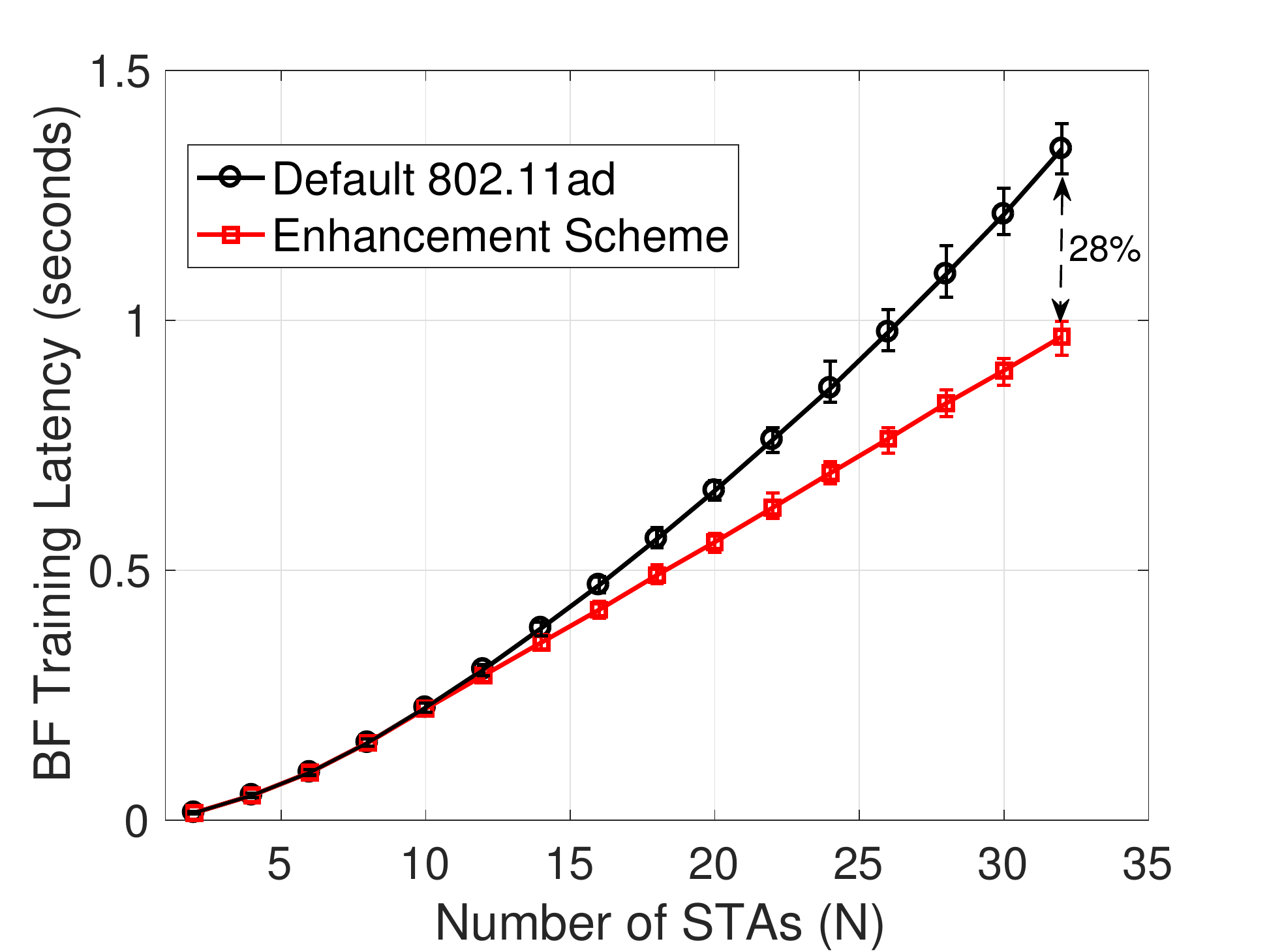}}
	\end{subfigure}%
	~
	\begin{subfigure}[M=12]{
			\label{fig:enhanced_delay_M=12}
			\includegraphics[width=0.4\textwidth]{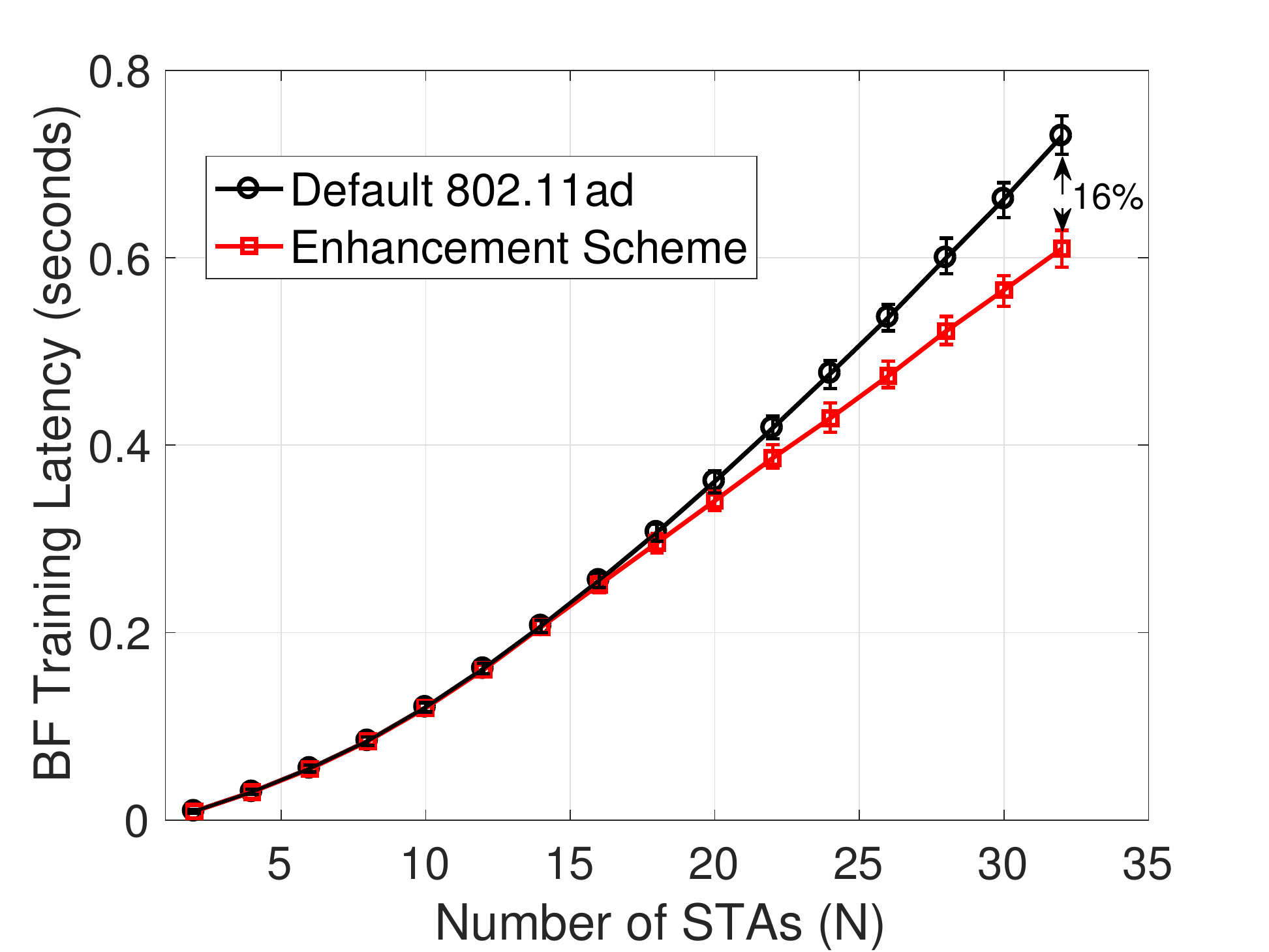}}
	\end{subfigure}
	\caption{Average BF training latency comparison in terms of different system parameters.}
	\label{Fig:enhanced_delayl}
	\vspace{-0.35cm}
\end{figure}

\subsection{Enhancement Scheme Evaluation}
In this subsection, we evaluate the performance of the proposed enhancement scheme by comparing with the default 802.11ad BF training protocol, in which both $R$ and $W$ adopt default configurations. 

We first evaluate the BF training efficiency of the proposed enhancement scheme in Fig. \ref{Fig:enhancedProtocol}. Several important observations can be obtained from simulation results. Firstly, the proposed enhancement scheme can significantly enhance the BF training efficiency in dense user scenarios as compared to the benchmark, which is presented in Fig. \ref{fig:enhancedProtocol_M=8}. Specifically, we show that a 35\% performance gain can be achieved when $N=32$. The key reason is that the proposed scheme can adjust protocol parameters in tune with user density to achieve the best performance. Secondly, %it is worth noting that two schemes achieve nearly the same performance in low user density scenarios. This is because the key difference between two schemes is the backoff mechanism which mainly works in dense user scenarios. %The maximum normalized throughput of these three schemes is nearly the same, which complies with the theoretical analysis in Section \ref{sec: maxium normalized throughput}. 
%Thirdly, 
the performance gap between two schemes narrows as the number of A-BFT slots increases. As reported in Fig. \ref{fig:enhancedProtocol_M=12} for $M=12$, the proposed enhancement scheme only achieves 17\% performance gain as compared to the default 802.11ad. Hence, simulation results show that the proposed enhancement scheme is more suitable for dense user scenarios with insufficient A-BFT slots.

We then present the average BF training latency comparison between the proposed enhancement scheme and the default 802.11ad protocol for $M=8, 12$, as shown in Fig. \ref{Fig:enhanced_delayl}. The average BF training latency of two schemes increases with the number of STAs because of the increase of the collision probability.  As compared to the default 802.11ad,  the proposed scheme can achieve a considerable latency reduction. Specifically, for a large number of STAs ($N=32$), a 28\% latency reduction can be observed clearly in Fig. \ref{fig:enhanced_delay_M=8}. Similar to the BF training efficiency, as the number of A-BFT slots grows, the performance gain on the average BF training latency drops, because the increase of A-BFT slots relieves the collision. As shown in Fig. \ref{fig:enhanced_delay_M=12}, for $M=12$, the proposed scheme still obtains a 16\% latency reduction when $N=32$, which further validates the effectiveness of the proposed scheme. %Although the BF training latency can be reduced via adjusting protocol parameters, the latency is still up to hundreds of milliseconds in dense user scenarios. %Low-latency yet efficient MAC protocols are still desired. 

\begin{figure}[t]
	\centering
	\renewcommand{\figurename}{Fig.}
	\includegraphics[width=0.4\textwidth]{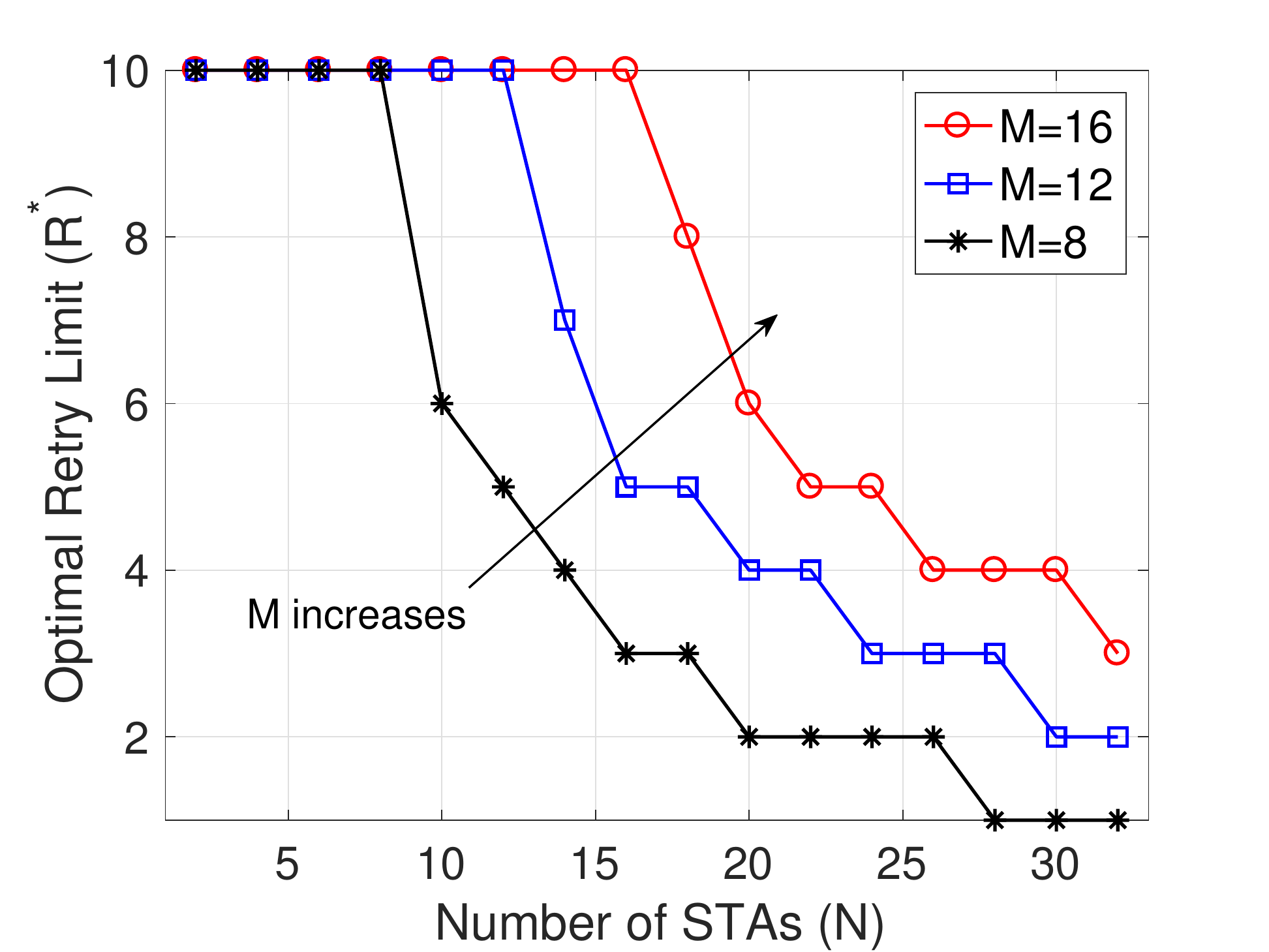}
	\caption{The optimal value of the retry limit in terms of the number of STAs.}
	\label{fig: optimal_retry_limit}
	\vspace{-0.35cm}
\end{figure}

Finally, Fig. \ref{fig: optimal_retry_limit} shows the optimal values of the retry limit with respect to the number of STAs for $M=8, 12, 16$. The optimal value of the retry limit decreases with the number of STAs. The simulation results show that a small value of the retry limit is preferred in dense user scenarios. For example, for $M=8$, the optimal value of the retry limit decreases to 1 when the number of STAs is larger than 28. The reason is that a small value of the retry limit in dense user scenarios renders STAs prone to enter backoff states. Then, the collision probability can be reduced to enhance the BF training efficiency. In addition, with the increase of A-BFT slots, the BF training procedure becomes less congested,  and hence a larger value of the retry limit should be chosen. For instance, when $N=32$, the optimal value of the retry limit is 3 for $M=16$, which is larger than that for $M=8$.

%-----------------------------------------------------------------------------------------------------------------------------------------------
\section{Conclusion and Future Work}\label{sec.Conclusion}
In this paper, we have investigated the performance of the BF training protocol in 802.11ad. An accurate analytical model has been developed to analyze the BF training efficiency and latency, which is validated by extensive simulation results. Theoretical analysis has unveiled the impacts of various protocol components, including the number of A-BFT slots, user density and the protocol parameters, on the BF training performance. %Based on the analytical model, approximate analysis has demonstrated that the maximum BF training efficiency is  $1/e$.  
To enhance the BF training performance in dense user scenarios, we have proposed an enhancement scheme which can effectively improve the BF training efficiency as well as reducing the BF training latency.  %For future work, due to the low throughput of the MAC protocol, we aim to design novel efficient MAC protocols.

Since 802.11ay is expected to adopt a similar BF training protocol, the proposed analytical model can be readily extended to future 802.11ay systems. This work can be viewed as an effective attempt towards the performance analysis of BF training protocol in mmWave networks. For our future work, we will investigate the protocol performance in mobile scenarios.

%Analytical results have been verified through discrete event simulations, which provide guidelines for optimal parameter setting in 802.11ad MAC in dense networks. 

%Through the theoretical analysis, we have demonstrated the inefficiency of 802.11ad MAC protocol for BF training.

\section{ACKNOWLEDGMENT}
This work was financially supported by Huawei Canada Co., Ltd. The authors thank Yan Xin,  Osama Aboul-Magd and Edward Au from Huawei Canada for valuable suggestions and comments. The authors thank Qinghua Shen, and Miao Wang (Miami University) for very helpful feedback and suggestions on the paper.

%--------------------------------------------------------------------------------------------------------------------------------------------------------------------------------------------------------------------
\appendix
\subsection{Proof of Theorem \ref{theorem_1}}\label{AppA}
Based on the proposed analytical model, the steady state probability vector can be solved with the following three steps. 

Firstly, with the one-step transition probability in \eqref{equ:STD1Prob}, we know that $\pi_{r+1,0}=p\cdot \pi_{r,0},\forall r\in [0,R-2]$. Hence, the steady state probability at state $(R-1,0)$ can be represented by
\begin{equation}\label{equ:probability_state_active}
\pi_{R-1,0}=p^{R-1}\pi_{0,0}.
\end{equation}

Secondly, with the one-step transition probability in backoff states \eqref{equ:STD3Prob}-\eqref{equ:STD4Prob0}, the steady probability for backoff states can be given as follows:
\begin{equation}\label{equ:probability_backoff_state}
\pi_{R,w}=\frac{\left(W-w\right)p}{W} \left(\pi_{R,0}+\pi_{R-1,0}\right), \forall w\in [0,W-1].
\end{equation}
Specifically, it can be obtained that $\pi_{R,0}=\frac{p}{1-p}\pi_{R-1,0}$ by taking $w=0$ in  \eqref{equ:probability_backoff_state}. Thus, according to \eqref{equ:probability_state_active},  \eqref{equ:probability_backoff_state} can be rewritten as
\begin{equation}\label{equ:probability_final}
\begin{split}
\pi_{R,w}
&=\frac{\left(W-w\right)p}{W\left(1-p\right)}\pi_{R-1,0}\\
&=\frac{\left(W-w\right) p^R}{W\left(1-p\right)}\pi_{0,0}, \forall w \in[0,W-1].
\end{split}
\end{equation}
%and 
%\begin{equation}
%\pi_{R,w}=
%\end{equation}

Finally, substituting \eqref{equ:probability_state_active} and \eqref{equ:probability_final} into \eqref{equ:balance2}, $\pi_{0,0}$ can be solved as
\begin{equation}
\pi_{0,0}=\frac{\left(1-p\right)}{p^R\left(W-1\right)/2+1}.
\end{equation}

Since other steady state probabilities can be represented by $\pi_{0,0}$, as shown in \eqref{equ:probability_state_active} and \eqref{equ:probability_final}, Theorem \ref{theorem_1} is proved.

%
%\begin{equation}
%\begin{split}
%{p}\approx1- e^{-{N\tau}/{M}}
%\end{split}
%\end{equation}

\bibliographystyle{IEEEtran}
\bibliography{security}
\vspace*{-1.5\baselineskip}
\begin{IEEEbiography}[{\includegraphics[width=1in,height=1.25in,clip,keepaspectratio]{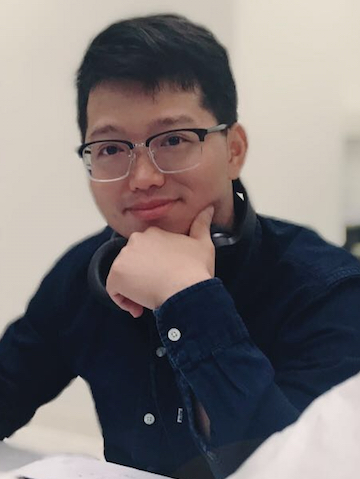}}]{Wen Wu}
	(S'13-M'20) earned the Ph.D. degree in Electrical and Computer Engineering from University of Waterloo, Waterloo, ON, Canada, in 2019. He received the B.E. degree in Information Engineering from South China University of Technology, Guangzhou, China, and the M.E. degree in Electrical Engineering from University of Science and Technology of China, Hefei, China, in 2012 and 2015, respectively. Starting from 2019, he works as a Post-doctoral fellow with the Department of Electrical and Computer Engineering, University of Waterloo. His research interests include millimeter-wave networks and AI-empowered wireless networks.
\end{IEEEbiography}

\vspace*{-1.5\baselineskip}
\begin{IEEEbiography}[{\includegraphics[width=1in,height=1.25in,clip,keepaspectratio]{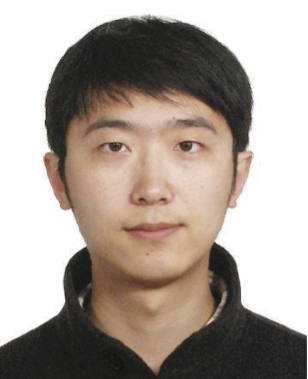}}]{Nan Cheng}
	(M'16) received the Ph.D. degree from the Department of Electrical and Computer Engineering, University of Waterloo in 2016, and B.E. degree and the M.S. degree from the Department of Electronics and Information Engineering, Tongji University, Shanghai, China, in 2009 and 2012, respectively. He is currently a professor with School of Telecommunication Engineering, and with State Key Lab of ISN, Xidian University, Shaanxi, China. He worked as a Post-doctoral fellow with the Department of Electrical and Computer Engineering, University of Toronto, from 2017 to 2018. His current research focuses on space-air-ground integrated system, big data in vehicular networks, and self-driving system. %His research interests also include performance analysis, MAC, opportunistic communication, and application of AI for vehicular networks.
\end{IEEEbiography}

\vspace*{-1.5\baselineskip}
\begin{IEEEbiography}[{\includegraphics[width=1in,height=1.25in,clip,keepaspectratio]{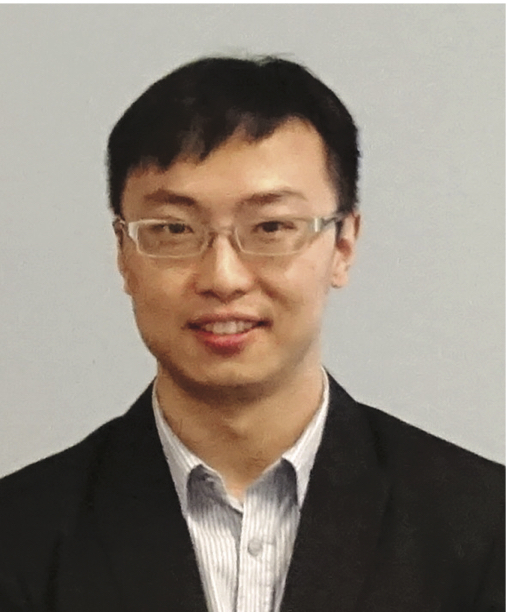}}]{Ning Zhang}
	(M'15-SM'18) received the Ph.D degree from University of Waterloo, Canada, in 2015. After that, he was a postdoc research fellow at University of Waterloo and University of Toronto, Canada, respectively. Since 2017, he has been an Assistant Professor at Texas A\&M University-Corpus Christi, USA. He serves as an Associate Editor of IEEE Internet of Things Journal, IEEE Transactions on Cognitive Communications and Networking, IEEE Access and IET Communications. His current research interests include wireless communication and networking, mobile edge computing, machine learning and physical layer security.	
\end{IEEEbiography}

\vspace*{-1.5\baselineskip}
\begin{IEEEbiography}[{\includegraphics[width=1in,height=1.25in,clip,keepaspectratio]{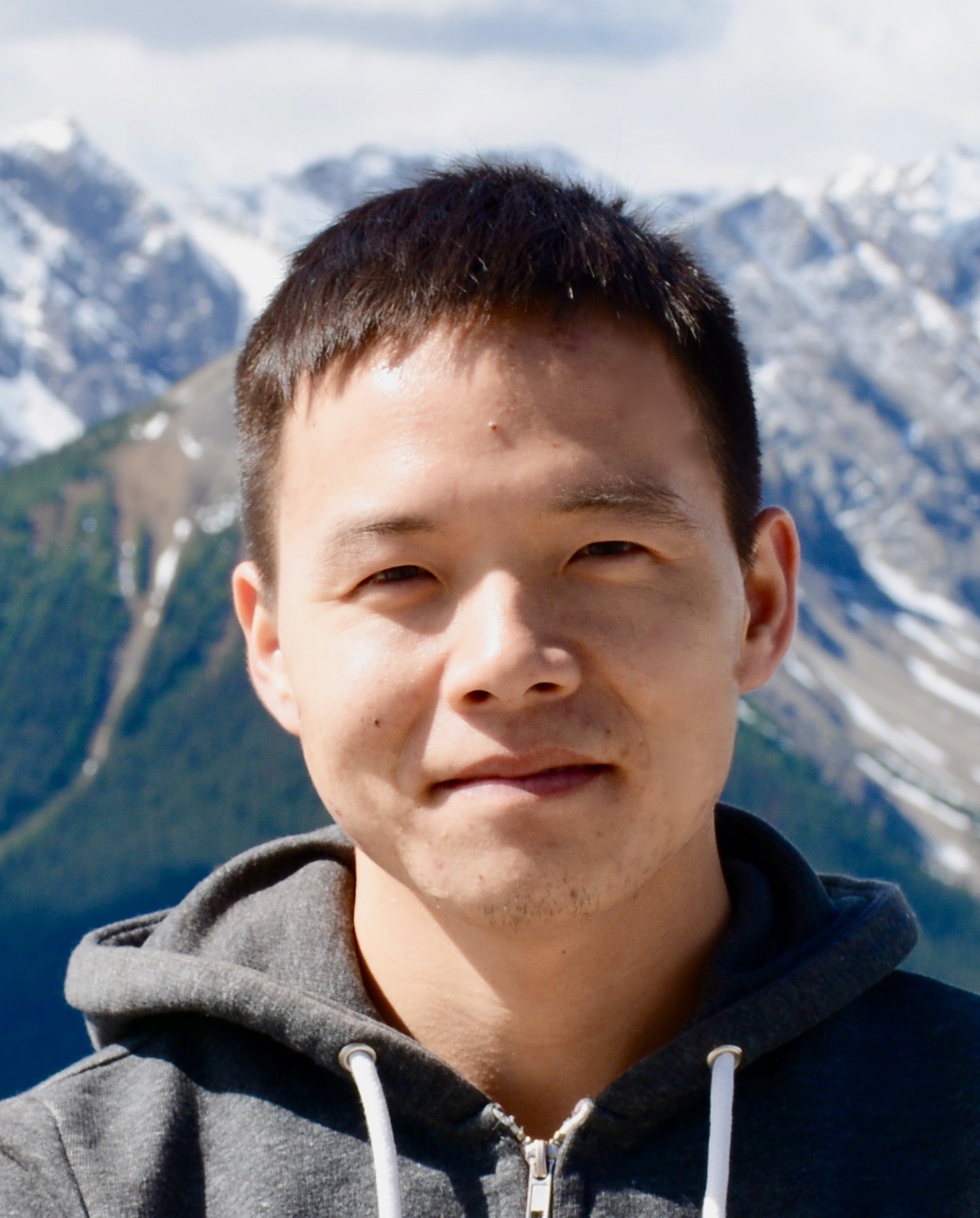}}]{Peng Yang}
	(S'16-M'18) received his B.E. degree in Communication Engineering and Ph.D. degree in Information and Communication Engineering from Huazhong University of Science and Technology (HUST), Wuhan, China, in 2013 and 2018, respectively. He was with the Department of Electrical and Computer Engineering, University of Waterloo, Canada, as a Visiting Ph.D. Student from Sept. 2015 to Sept. 2017, and a Postdoctoral Fellow from Sept. 2018 to Dec. 2019. Since Jan. 2020, he has been a faculty member with the School of Electronic Information and Communications, HUST. His current research focuses on mobile edge computing, video streaming and analytics.
\end{IEEEbiography}

\begin{IEEEbiography}[{\includegraphics[width=1in,height=1.25in,clip,keepaspectratio]{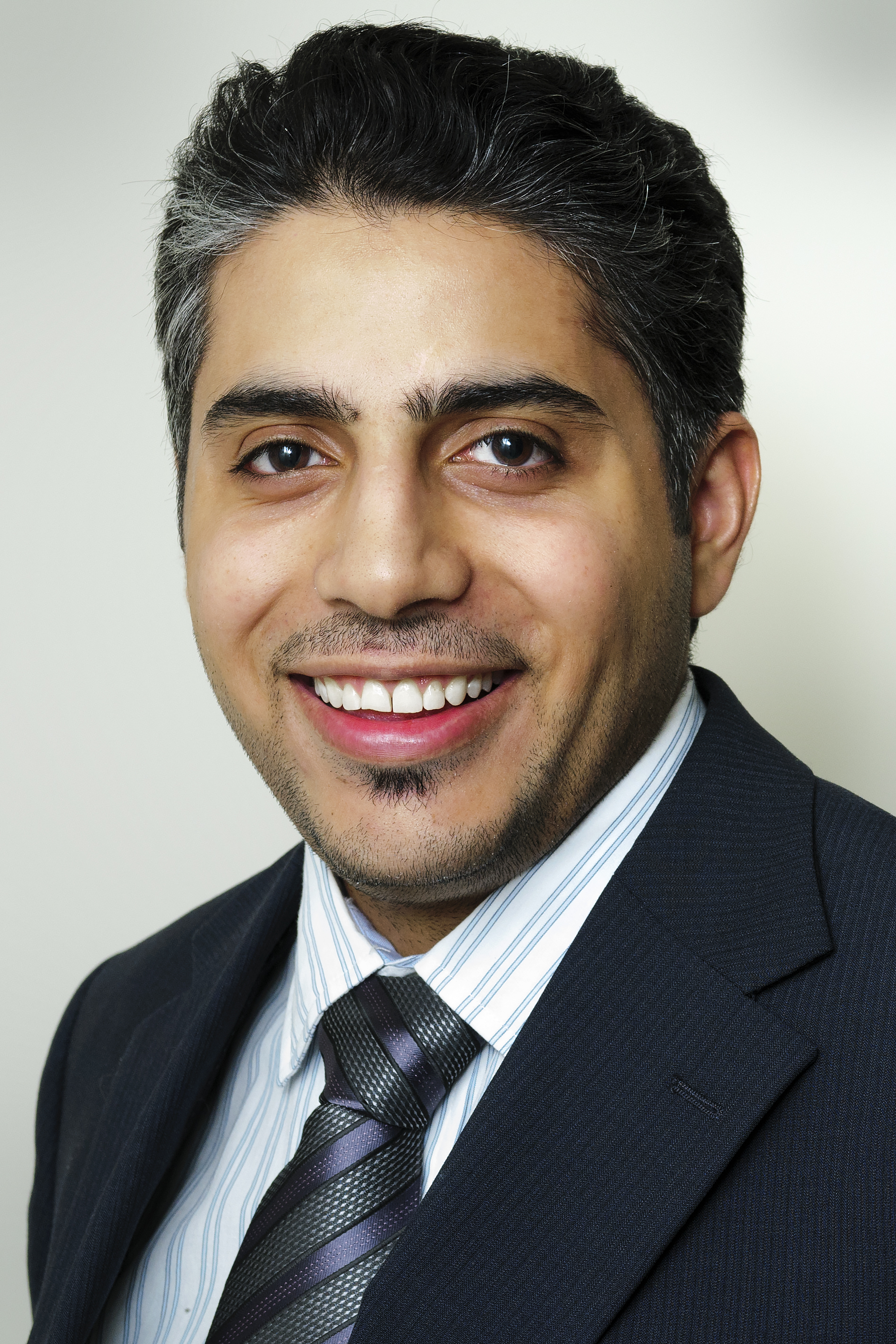}}]{Khalid~Aldubaikhy}
	(S'10) is currently an Assistant Professor at Qassim University, Buraydah, Al-Qassim, Saudi Arabia. He received the B.E. degree from Qassim University, Saudi Arabia, in 2008, the M.A.Sc. degree in Electrical and Computer Engineering from Dalhousie University, Halifax, NS, Canada, in 2012, and the Ph.D. degree in Electrical and Computer Engineering from University of Waterloo, Waterloo, ON, Canada, in 2019. His research interests include millimeter-wave wireless	networks, medium access control, and mmwave 5G networks.
\end{IEEEbiography}

\vspace*{-1.5\baselineskip}
\begin{IEEEbiography}[{\includegraphics[width=1in,height=1.25in,clip,keepaspectratio]{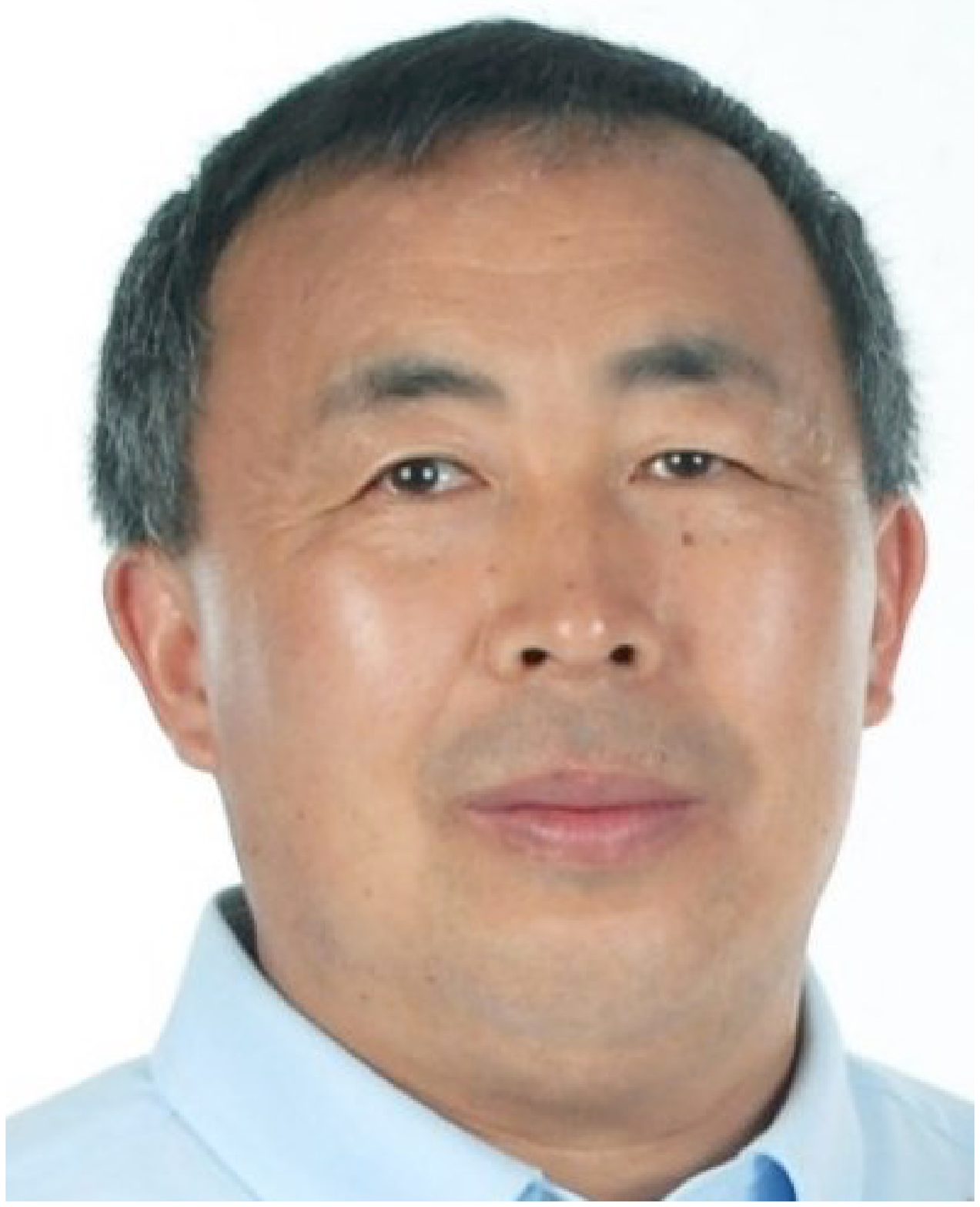}}]{Xuemin (Sherman) Shen}(M'97-SM'02-F'09) received the Ph.D. degree in electrical engineering from Rutgers University, New Brunswick, NJ, USA, in 1990. He is currently a University Professor with the Department of Electrical and Computer Engineering, University of Waterloo, Canada. His research focuses on network resource management, wireless network security, Internet of Things, 5G and beyond, and vehicular ad hoc and sensor networks. He is a registered Professional Engineer of Ontario, Canada, an Engineering Institute of Canada Fellow, a Canadian Academy of Engineering Fellow, a Royal Society of Canada Fellow, a Chinese Academy of Engineering Foreign Fellow, and a Distinguished Lecturer of the IEEE Vehicular Technology Society and Communications Society. 

Dr. Shen received the R.A. Fessenden Award in 2019 from IEEE, Canada, James Evans Avant Garde Award in 2018 from the IEEE Vehicular Technology Society, Joseph LoCicero Award in 2015 and Education Award in 2017 from the IEEE Communications Society. He has also received the Excellent Graduate Supervision Award in 2006 and Outstanding Performance Award 5 times from the University of Waterloo and the Premier’s Research Excellence Award (PREA) in 2003 from the Province of Ontario, Canada. He served as the Technical Program Committee Chair/Co-Chair for the IEEE Globecom’16, the IEEE Infocom’14, the IEEE VTC’10 Fall, the IEEE Globecom’07, the Symposia Chair for the IEEE ICC’10, and the Chair for the IEEE Communications Society Technical Committee on Wireless Communications. He was the Editor-in-Chief of the IEEE INTERNET OF THINGS JOURNAL and IEEE Network, and the Vice President on Publications of the IEEE Communications Society.
\end{IEEEbiography}

\end{document}